\documentclass[twocolumn,aps,preprintnumbers,amsmath,amssymb]{revtex4}
\usepackage[]{graphicx}
\usepackage{xcolor}

\usepackage{adjustbox}
\usepackage{array}
\usepackage{soul}
\newcolumntype{C}{>{\centering\arraybackslash\large}m{1.6cm}}
\newcolumntype{L}{>{\centering\arraybackslash\large}m{3.2cm}}
\newcolumntype{M}{>{\centering\arraybackslash\large}m{3.5cm}}

\newcommand{\ian}[1]{\textcolor{cyan}{[\emph{#1}]}}
\newcommand{\wolf}[1]{\textcolor{blue}{[\emph{#1}]}}

\newcommand{\md}{\mbox{d}}
\newcommand{\trans}{^{\rm T}}
\newcommand{\inv}{^{-1}}
\newcommand{\uv}{\mathbf{u}}
\newcommand{\xv}{\mathbf{x}}
\newcommand{\zv}{\mathbf{z}}
\newcommand{\rv}{\mathbf{r}}
\newcommand{\ev}{\mathbf{e}}
\newcommand{\cv}{\mathbf{c}}
\newcommand{\Amat}{\mathbf{A}}
\newcommand{\Bmat}{\mathbf{B}}
\newcommand{\Emat}{\mathbf{E}}
\newcommand{\Imat}{\mathbf{I}}
\newcommand{\Qmat}{\mathbf{Q}}
\newcommand{\Rmat}{\mathbf{R}}
\newcommand{\Mmat}{\mathbf{M}}
\newcommand{\Cmat}{\mathbf{C}}

\usepackage[outdir=./]{epstopdf}

\begin{document}

\title{Predicting synchronized gene coexpression patterns from
  fibration symmetries in gene regulatory networks in bacteria}

\author{Ian Leifer}
\affiliation{Levich Institute, Physics Department, City College
  of New York, New York, NY 10031}
\author{Mishael S\'anchez-P\'erez}
\affiliation{Levich Institute and Physics Department, City College
  of New York, New York, NY 10031}
\affiliation{Programa de Gen\'omica Computacional,
Centro de Ciencias Gen\'omicas,
Universidad Nacional Aut\'onoma de M\'exico, Cuernavaca, Mexico}
\author{Cecilia Ishida}
\affiliation{Faculty of Medicine and Biomedical Sciences. Autonomous University of Chihuahua, 31125 Chihuahua, Chihuahua, M\'exico}
\author{Hern\'an A. Makse \footnote{Correspondence to hmakse@ccny.cuny.edu}}
\affiliation{Levich Institute and Physics Department, City College
  of New York, New York, NY 10031}

\begin{abstract}

Background: Gene regulatory networks coordinate the expression of
genes across physiological states and ensure a synchronized expression
of genes in cellular subsystems, critical for the coherent functioning
of cells.  Here we address the questions whether it is possible to
predict gene synchronization from network structure alone.  We have
recently shown that synchronized gene expression may be predicted from
symmetries in the gene regulatory networks (GRN) and
described by the concept of symmetry fibrations.  We showed that
symmetry fibrations partition the genes into groups called fibers
based on the symmetries of their 'input trees', the set of paths in
the network through which signals can reach a gene.  In idealized
dynamic gene expression models, all genes in a fiber are perfectly
synchronized, while less idealized models -- with gene input functions
differencing between genes -- predict symmetry breaking and
desynchronization.

Results: To study the functional role of gene fibers and to test
whether some of the fiber-induced coexpression remains in reality, we
analyze gene fibrations for the gene regulatory networks of
\emph{E.~coli} and \emph{B.~subtilis} and confront them with
expression data.  We find approximate gene coexpression patterns
consistent with symmetry fibrations with idealized gene expression
dynamics. This shows that network structure alone provides useful
information about gene synchronization, and suggest that gene input
functions within fibers may be further streamlined by evolutionary
pressures to realize a coexpression of genes.

Conclusions: Thus, gene fibrations provides a sound conceptual tool to
describe tunable coexpression induced by network topology and shaped
by mechanistic details of gene expression.

\end{abstract}

\maketitle

\section{Introduction}


Gene regulation in bacteria has been studied since the time of the
operon model of Jacob and Monod \cite{monod}.  Knowing the regulators
and mechanistic details of genes expression has greatly improved our
understanding of cellular signal processing \cite{klipp}, and has lead
to various applications in systems and synthetic biology
\cite{klipp,palsson,gerosa,karlebach}.  General factors like RNA
polymerase activity can lead to an overall increase or decrease of
bacterial gene expression depending on cellular growth rate.  The
expression profiles of single genes are further regulated by specific
transcription factors (TFs).  At the same time, high-throughput
expression studies have revealed the functional role of expression
profiles: a variety of multivariate methods (including clustering,
biclustering, plaid models, singular value decomposition, and
Independent Component Analysis) have been used to extract functional
information from expression profiles, often taking co-expression as a
sign for shared biological function.

Gene regulation by transcription factors is described by
gene regulatory networks (GRN)
\cite{caldarelli,karlebach,young} where nodes are genes and a directed
edge from gene A to gene B states that gene product of A is a TF that
regulates the expression of B as an activator, repressor, or dual
regulator. TF activities may be further modulated by signaling
molecules that bind to the TF to activate or inactivate the
protein. This process conveys information about the state of the cell,
implementing for example a negative feedback control from metabolic
synthesis pathways. In the gene regulatory network, such effector
signaling molecules appear as external inputs to the GRN.  The
topologies of GRNs have been studied in detail \cite{caldarelli} and
been used as blueprints for dynamic models of gene expression
\cite{karlebach}.  Such models describe the production and degradation
of gene products (mRNA or proteins) and consider the regulatory input
functions of individual genes, which reflect TF binding to genes'
promoter regions, with gene-dependent binding parameters and possible
binding states.

In principle, synchronized activity in gene expression
\cite{arenas,cecilia1,cecilia2,clocks,hasty,stricker,tigges} in
bacteria reflects the gene arrangement in operons since genes in
operons are transcribed together into a single mRNA molecule. A
transcriptional unit (TU) is a set of contiguous genes that are
transcribed into one mRNA. An operon is a set of contiguous genes
controlled by the same promoter.  However, beyond this trivial
synchronization in gene expression, the largest part of the
co-expression synchronization of gene activity can actually be
attributed to specific pathways in the GRN regulated by specific TFs.



Quantitative gene expression models based on realistic input functions
for all of the genes are out of reach due to the multiplicity of
parameters defining these input functions \cite{alon}. Thus, there
have been attempts to understand dynamic properties from network
structure alone \cite{alon,smma:02,majt:08,gbmn:08}. To establish GRNs
for major model organisms, TF binding has been predicted from binding
site sequences and based on ChIP/chip, ChIP-seq or ChIP-exo
experiments.  In these networks, motifs like the feed-forward loop
\cite{alon,alon-motif,smma:02} have been found by statistical methods,
based on the frequent occurrence in networks, and have been studied as
local signal processing circuits embedded in larger networks.

In a series of recent papers, we showed that symmetry principles
applied to biology networks \cite{connectomeMorone,pnas,ian} can
explain the synchronization in gene coexpression profiles.
\textcolor{red}{Synchronization here denotes joint evolution in time. That is, if each gene $i$ in the network is represented by it's expression level $x_i(t)$, as measured by the
protein gene product with the time-scale approximations made,
typically the concentration of mRNA in the cell, then
genes $i$ and $j$ with expression levels $x_i$ and $x_j$ are synchronized if $lim_{t \rightarrow \infty}(x_i(t)-x_j(t)) \rightarrow 0$.}
We have
found structural symmetries in biological networks, described by the
concept of symmetry fibrations \cite{pnas}, that provide a principled
way to define building blocks of genetic networks supporting
synchronized gene expression.  Symmetry fibrations are a powerful tool
to describe the structure of networks that combines geometry and
algebra data via category theory \cite{grothendieck,golubitsky,vigna}.

In this paper we first review the concept of symmetry fibration
applied to GRNs in bacteria and we further address the question of
gene synchronization and coexpression by comparing to existing
experimental data, and ask how gene coexpression can be robustly
implemented by network structure. \textcolor{red}{The methodology allows the user to find groups of genes that
are active at the same time. These groups may define trap spaces (groups of related
attractors, with their corresponding trap spaces) or even specific attractors especially
if the pattern of synchronous genes includes inactive genes as well as active ones. In this paper we address the question of how the behavior of the system relates to the synchronous gene groups.} In particular, we challenge the
predominant view that \emph{coexpression} (by which two genes show
similar expression profiles) is necessarily a sign of
\emph{coregulation} (by which these genes are controlled by a common
transcription factor)
\cite{wgcna,zhang-horvath,r-package,collins,wisdom,bansal,brugere,atul,hartung,chen,wolf-network,nca,roy,klipp,kaplan}. Instead,
we claim that other, more complex circuits in the regulatory network
can lead to robust coexpression. These circuits are identified by
their symmetry properties and show synchronization in gene
coexpression as a result of the underlying symmetries in the gene
regulatory network.


In the case of GRNs, fibration symmetries can reveal meaningful
building blocks in regulatory networks \cite{pnas,ian}. In contrast
to network motifs, which can only be found by their frequent
occurrence, gene fibrations can find such circuits based on their
symmetry properties even if these circuits are large and appear only
once in the network.
\textcolor{red}{Furthermore, fibrations are mathematically proven to induce synchronization, which will be discussed in more detail in Chapter~\ref{Chapter:Fibrations}.}
The symmetry fibration is a transformation
that reduces the networks to its base by collapsing symmetric genes
into fibers.  Genes that belong to one fiber are predicted to show
synchronous expression activity, resulting in coexpression patterns of
all genes in the fiber. These circuits can perform signal processing
tasks, namely generating structurally encoded, yet tunable patterns of
gene coexpression. Other types of building blocks, such as densely
connected modules defined by counting a 'density' of edges within that
module \cite{newman}, are not expected to show this property.

Thus, our previous work \cite{pnas,ian,ian-theory} shows that symmetry
fibrations can group genes into fibers that point us to the
synchronized building blocks of the GRN. Using the same approach, we
managed to detect regulatory structures in the neural network of
\emph{C.~elegans} worms and to give them a functional explanation
\cite{connectomeMorone}.  If fibers can explain synchronization
between cells, we can expect to see the same principle within cells,
in the coexpression of genes and the underlying gene regulatory
networks. Thus, as a working hypothesis, we pose that coexpression,
arising from transcriptional regulation, can be directly understood
through genome organization, gene conservation across genomes, and
network structure, namely through gene regulatory building blocks
related to fibers, while quantitative details of gene regulation play
a minor role.

To treat networks as ``blueprints'' of dynamic gene regulation models,
gene regulatory input functions must be defined.  With simple
identical input functions, and disregarding all quantitative
differences (e.g.~in binding parameters or mRNA half lives defining
the input functions), genes in a fiber are predicted to show identical
expression profiles.  Under this hypothesis, fibration symmetries lead
to important consequences for the dynamics of the gene expression in
the network: fibrations give rise to the existence of robustly
synchronous solutions.  In reality, however, input functions will
differ between genes, and additional modulation of transcription
factor activities by signaling molecules will come into play. The
predicted symmetry will be broken and we expect a partial
desynchronization. \textcolor{red}{Genes $i$ and $j$ with dynamical solutions $x_i(t)$ and $x_j(t)$ that reach the state in which after the time $t_{sync}$ solutions are $\varepsilon$-close are ``desynchronized''.
That is, two genes are almost synchronized or partially ``desynchronized'' if $(x_i(t)-x_j(t)) < \varepsilon$ for $t \in [t_{sync}, \infty]$. It was shown \cite{sorrentino2016b} that slight mismatch in the parameters leads to solutions with nearly synchronous trajectories, that is, desynchronized solutions. Detailed analysis of the situation with the difference in the input functions hasn't been done in the literature so far, but basing on the conclusion in the networks with slight mismatch, we assume that in the networks with ``slightly bigger'' mismatch perfect synchronization will be broken even further creating more desynchronization.}
We hypothesize that precise coexpression,
as predicted by our idealized model, will be reflected in cells in a
partial coexpression that may be tuned by external biochemical
signals.

\textcolor{red}{
Each gene (node of the network) and it's time evolution is thought of as one variable - it's expression level. Thus, it may look like this method is not applicable to more complex organisms because processes like transcription, translation, folding and binding to DNA are lumped into one step and sophisticated effects like RNA splicing, DNA methylation, histone acetylation, and overwinding or underwinding of DNA are ignored. However, fibration theory can still be applied even when one node has a very complicated behavior that takes all of the above into consideration. For example, we can extend each node to two variables - mRNA and protein concentration, and then the dynamics of the symmetric nodes will be synchronous by variable. That is, in this example synchronous nodes will have both equal mRNA and protein concentrations. As discussed above, each level of complexity has a potential to create more symmetry breaking. Thus, whether gene fibrations are useful in practice depends
on how much of the coexpression remains in reality.}
In this paper we study this question in detail by studying
coexpression in \emph{Escherichia coli} and \emph{Bacillus subtilis},
the major Gram-positive and Gram-negative bacterial model organisms: we predict set of coexpressed genes based on fibration symmetry and
confront these predictions with transcriptome data. The GRNs under
study, which we analysed before in \cite{pnas,ian}, contain various
types of fibers, including feed-forward fibers, Fibonacci fibers,
multilayer composite fibers, and $n=2$ fibers \cite{pnas}.
After
giving an overview of the fibers from \cite{pnas} in Section \ref{sec:hierarchy}, we
study expression of groups of genes predicted to be coexpressed in
Section \ref{results}.

The main approximation behind the existence of synchronization in
fibers is the 'uniformity assumption': the assumption that all the
parameters defining the input functions (i.e., the Hill function
defining the interaction of the TF with the binding site of the target
gene) of the genes in the fibers are the same.  That is, the Hill
input functions, reflecting interactions of TF binding to
DNA, as well as binding parameters or mRNA half lives, needs to be the
same for all genes in a fiber to synchronize. Thus, our prediction of
perfect synchronization in the fiber depends on a idealized model of
gene regulation, where the input functions of genes in the
fiber are the same. Under these conditions, fibration theory predicts
perfect synchronization in the fiber. Of course, in reality, input
functions and interaction parameters of different genes will not be
exactly the same, and the question becomes of how much
desynchronization is caused by this symmetry breaking in the input
functions and parameters of interactions.

Despite all the likely reasons for desynchronization ignored in our
working hypothesis, including differing gene input functions and
various levels of gene regulation, we find a measurable degree of
synchronization, matching our simplified theory. We conclude that
despite quantitative differences between gene input functions, genes
within fibers have measurably more synchronization, which justifies
our topological analysis in this case.
A further possibility, supported by previously measured gene input
functions, is that evolutionary pressures may be at work which
``streamline'' the gene input functions within fibers, thus preventing
a stronger symmetry breaking.

\section{The fibration formalism}
\label{Chapter:Fibrations}

The main concepts from graph fibration theory for biological networks
-- isomorphic input trees, fibers, symmetry fibrations, and the base
of a network -- have been introduced in Morone, Leifer \& Makse
\cite{pnas} (for details, see Methods section \ref{sec:Fibrations})
and are based on previous developments by Golubitsky and Stewart
\cite{golubitsky} and Boldi and Vigna \cite{vigna}. Example of symmetry
fibrations are displayed Figs.~\ref{regulon}-\ref{ff}.  All these
concepts will be exemplified below.  In the fibration formalism, a
network is described as a directed graph.  The input tree of a node
represents all paths in the network that lead to this node \cite{pnas}
and it exemplified in Fig.~\ref{ar} further below. An input tree is
constructed by considering the node of interest, which forms the root
of the tree, that is, the end point of all the paths leading to that
node.  If a path contains loops, these loops become ``unfolded'' in
the tree representation.  Then, every node in a given layer of the
tree represents the initial point of a path in the network leading to
the root node.  The first layer of the input tree contains all the
nodes that are connected by a direct arrow to the root nodes (that is,
by a path of length 1). The second layer contains nodes with paths of
length 2, and so on.  Two trees are called isomorphic if they are
topologically identical, where the identity of nodes in the tree does
not matter.  Nodes with isomorphic input trees are considered
equivalent and belong to the same fiber, as exemplified further below
in Figs.~\ref{regulon}-\ref{ff}.  A symmetry fibration of a graph $G$
\cite{pnas}, called a surjective minimal graph fibration in
\cite{vigna,deville,sanders}, is a transformation $\psi: G \to B$ that
collapses the nodes in each fiber of $G$ into a single representative
node, thus reducing the network $G$ to a graph $B$ called its base
(see Figs.~\ref{regulon}-\ref{ff}). In this way, a symmetry fibration
reduces a network to its most simple form by compressing the
redundancies provided by the symmetric genes in the fibers.


In a GRN -- with nodes representing genes or gene products --
fibration symmetries have important consequences for gene expression
dynamics.
\textcolor{red}{Theory developed in \cite{vigna, golubitsky, deville, sanders} shows that if there is a dynamical system that corresponds to the coupled-cell network (in this case the expression dynamics of the gene regulatory network), then the dynamics of the parts of the network that correspond to the different fibers can take a robustly synchronous state. That is, for the network $G$ with fibers $f_1, f_2, \dots f_n$ consisting of genes $i_1, i_2, \dots i_{\mid f_i \mid}$ for $i \in \{1 \dots n\}$, where $\mid f_i \mid$ is the size of fiber $i$, there exists a solution that is robustly synchronous by fiber i.e.}

\begin{equation}
\color{red}
  \begin{cases}
    x_{1_1}(t) = x_{1_2}(t)=\dots=x_{1_{\mid f_1 \mid}}(t), \\
    x_{2_1}(t) = x_{2_2}(t)=\dots=x_{2_{\mid f_2 \mid}}(t), \\
    \dots \\
    x_{n_1}(t) = x_{n_2}(t)=\dots=x_{n_{\mid f_n \mid}}(t).
  \end{cases}
\end{equation}

\textcolor{red}{Depending on the type of coupling and parameter values this solution can be stable or unstable. Additionally, stable solutions can have different size of the basin of attraction. Numerical solutions of the dynamical evolution of the fibers studied here and performed in \cite{ian} indicate that for the particular type of interaction fibers found in genetic networks are stable and have a very big basin of attraction. Analytical investigation of the stability and attractiveness of these solutions is out of the scope of this paper, but it can be done using the approach introduced by Pecora, Sorrentino and collaborators \cite{pecora1, pecora2, sorrentino2019, sorrentino2020}.}

Since genes in fibers are synchronized and therefore redundant from
the point of view of the dynamics, the symmetry fibration reduces the
network to a smaller number of nodes, while preserving the dynamical
state of the network. As discussed, this
statement relies on a strong uniformity assumption about Hill input
functions, which will be scrutinized in subsequent sections.

To summarize, fibration is a mapping between graphs that collapses
fibers. Fibers are nodes that have isomorphic input trees and can be
collapsed together by map (fibration) with these
properties. Fibrations symmetries is a term we use to describe
symmetries that we are using.  The fibers
represent the functional set of synchronized genes in the GRN.  For
the GRNs of {\it E. coli} and {\it B.~subtilis} and other species, we
have previously organized the different types of observed fibers into
a hierarchy, reflecting the different complexity of the topological
features of their input trees \cite{pnas,ian}.
\textcolor{red}{This hierarchy identifies a broad range of fibers: 91 in \emph{E.~coli} (see \cite{pnas} for the full list of fibers and Table~\ref{Table:SignificanceNoFilter} for the distribution of the 85 fiber sizes between 2 and 24) and 216 in \emph{B.~subtilis} (203 fibers of sizes 2 to 24 are considered and their size distributed is shown in the Table~\ref{tbl:distributionSubtilis}).}
Despite their various topologies, the
resulting fibers can be concisely classified with just two numbers,
called 'fiber numbers' $\rvert  n, \ell \rangle$, which specify the number
of loops in the fiber, $n$, and of external regulators of the fiber
$\ell$ \cite{pnas}.

\textcolor{red}{The fibrations of GRNs can be computed by algorithms to find 'minimal balanced coloring' available in the literature \cite{kamei,pnas}. Equivalence between fibers and minimal balanced coloring is discussed in Methods Section \ref{sec:EquivalenceFibers} and a more detailed description of the algorithm is given in Methods Section \ref{sec:AlgorithmBalancedColoring}.}
Code to find
fibers is available at
\url{https://github.com/ianleifer/fibrationSymmetries}. Mathematical work on
fibrations \cite{golubitsky,vigna} has been concerned with the
existence of synchronized solutions, but in some cases realization of
the synchronized solution is quite unprobable.
We can imagine a
network with just 2 nodes with no edges. These two nodes can be
synchronized since synchronous solution exists for the two nodes, yet,
these solutions may not be most probable (small basin of attraction)
since they require some ad-hoc set of initial conditions. In terms of
the mathematical definition of symmetry fibrations, the existence of
these synchronous solutions is consistent.  However, in the real
system we obviously do not require these two nodes to belong to the
same fiber. We solve this problem in the following way. We first find
all strongly connected components with no input (including single
nodes with no input) and assign different then to different fibers.
In this way we avoid finding these stray solutions.

\section{Gene regulatory networks of \emph{E.~coli} and
  \emph{B.~subtilis}}

In this paper, we first review the fibers found in the GRNs of
\textit{E.~coli} and \textit{B.~subtilis} studied in \cite{pnas,ian}
and then organize this rich set of fibers into a well-defined
hierarchy.
At the most simple and trivial level in this hierarchy, we
find the known structures of operons and regulons that trivially lead
to synchronization.  The hierarchy then builds up to more complex
architectures as the fiber numbers $\rvert n, \ell \rangle$, describing
topological features of the fibers such as $n$ loops and $\ell$
regulators, increase, and progresses to fibers like the autoregulation
loops, feed-forward fibers, Fibonacci fibers, $n=2$ fibers and
multilayer composite fibers.

\begin{table}[h]
  \begin{tabular}{| >{\bfseries}>{\color{red}}c | >{\color{red}}c | >{\bfseries}>{\color{red}}c | >{\color{red}}c | >{\bfseries}>{\color{red}}c | >{\color{red}}c | >{\bfseries}>{\color{red}}c | >{\color{red}}c |}
   \hline
   Size & Count & Size & Count & Size & Count & Size & Count\\ 
   \hline
    2 &  65 & 7  & 8 & 12 & 5 & 21 & 2 \\ 
    3 &  40 & 8  & 2 & 13 & 1 & 22 & 1\\ 
    4 &  27 & 9  & 4 & 15 & 2 & 23 & 1\\ 
    5 &  16 & 10 & 6 & 16 & 2 & 24 & 1\\ 
    6 &  12 & 11 & 6 & 18 & 2 &    &  \\ 
  \hline
  \end{tabular}
  \caption{Distribution of the fiber sizes of \textit{B.~subtilis} for sizes between 2 and 24.}
  \label{tbl:distributionSubtilis}
\end{table}

The studied networks are constructed from data in RegulonDB for {\it
  E. coli} \cite{regulondb} and Subtiwiki \cite{SubtiWiki} for {\it
  B. subtilis}, well curated resources for gene annotations,
regulation, and function in the two bacteria. From the TF-gene
bipartite networks we construct a GRN with directed weighted
links. Nodes of these networks represent genes and edges are
interactions between source gene producing TF (source) that binds to
the DNA sequence upstream to the other gene (target). Edges are
considered as three different types according to their function:
activation, repression, and dual.

The regulons are the first (trivial) members in this hierarchy with a
structure represented by no loops and $\ell$ regulators: $\rvert n=0,\ell
\rangle$. Next, the scheme classifies non-trivial fibers by $\rvert 
n=1,\ell \rangle$ as feed-forward fibers (FFF) and autorregulation
loops (AR), and $\rvert  n=1,\ell \rangle$ as binary tree fibers. The list
is completed with more complex fibers with non-integers fiber numbers
called Fibonacci fibers with $\rvert \phi_d, \ell \rangle$, where $\phi_d$
is the generalized golden ratio, and composite multilayer fibers as
combinations of the fibers above.  We have shown in \cite{ian} that
this set of fibers arises as a constructive procedure that mimics a
growth procedure by recursively iterating a constructive process that
expands the existence of all fibers in the network.
We elaborate on this hierarchy in the rest of this
section. In Section \ref{results}, we shall test the biological
significance of these fibers by testing the prediction of
synchronization inside the fibers.

\section{Hierarchy of symmetry fibers in GRN}
\label{sec:hierarchy}

An important concept in this work is the distinction between
coexpression resulting from trivially sharing the same operon and
regulon versus synchronization induced by more complex symmetry
fibration. That is, the difference between coregulation by fiber
(coexpression resulting from shared input trees which take into
account extended paths in the network) vs. coregulation by a single
input of a regulon. Both of them lead to coexpression, but the former
is more complex than the latter, which is considered trivial
synchronization, while the fiber synchronization is not.  We
ellaborate on these distinction in the next two subsections.

\subsection{Operons and regulons}

The trivial building blocks leading to synchronization are operons and
regulons. Operons are gene arrangements ubiquitous in bacteria
\cite{monod}: genes in an operon have a common promoter and are not
transcribed into individual mRNAs, but as transcription units,
yielding a single mRNA strand containing several contiguous
genes. Depending on the locations of promoters and terminators, the
transcription units can also be overlapping.  The expression of genes
in an operon will be automatically synchronized since they are
translated together. In the case of multi-promoter operons, we can
group the genes into minimal transcription units, each being
controlled by the same combination of promoters. The genes in such a
minimal units should be precisely coexpressed (operons with several
terminators can be subdivided similarly). In our analysis we take the
operon as a single node in the GRN. When two or more TFs belong to the
operon, we leave one TF associated with the operon and separate the
remaining TF from the operon.


\subsection{Regulons}

\begin{figure}
  \centering
  \includegraphics[width=\linewidth]{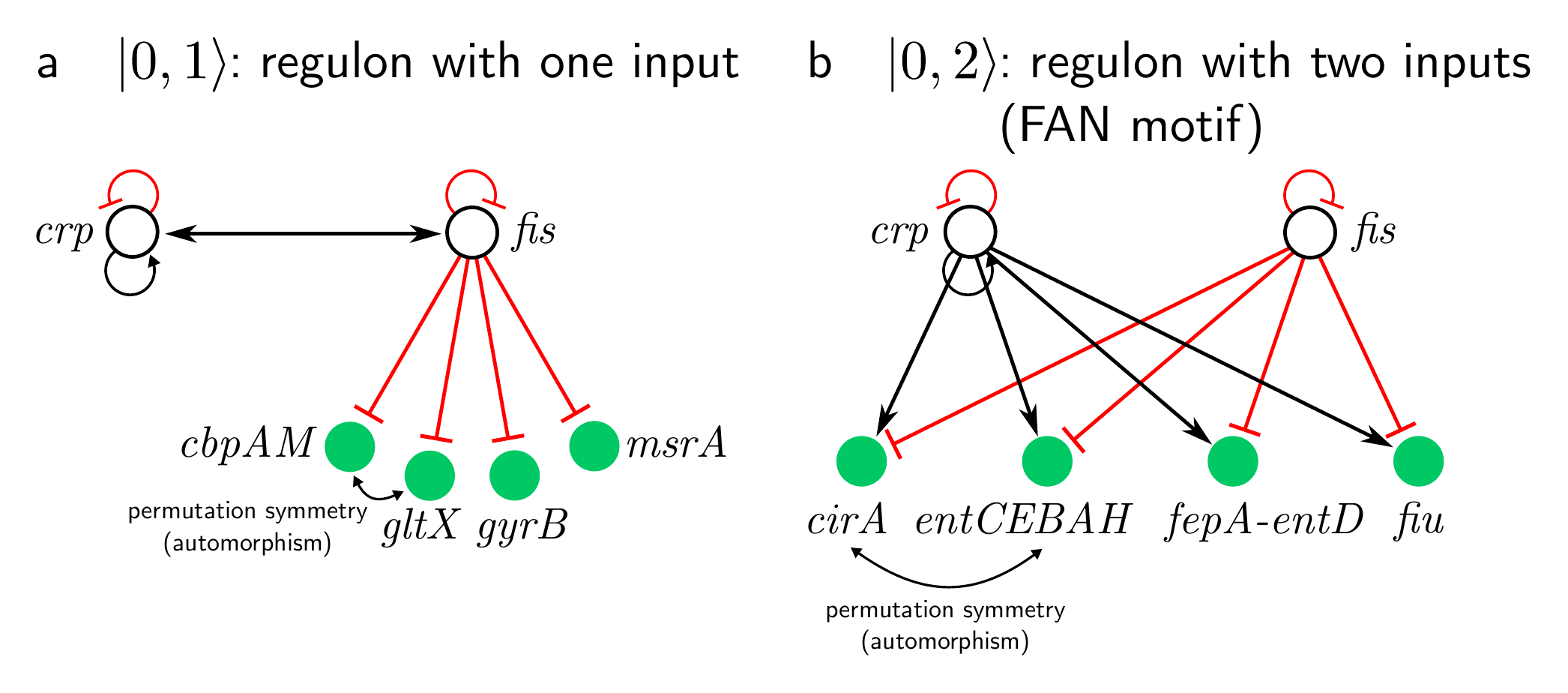}
  \caption{Trivial circuits leading to synchronization: Regulons with co-regulation. \textbf{a)} Genes \textit{cbpAM, gltX, gyrB} and \textit{msrA} are controlled by the same TF (\textit{fis}). Fiber numbers describing this circuit are $\rvert n = 0, \ell = 1 \rangle$ since there are no loops and fiber has 1 regulator. Gene activity can synchronize, because any two nodes can be permuted without the change in the network under the $S_4$ symmetry group. \textit{fis} won't be synchronized with \textit{cbpAM, gltX, gyrB, msrA}, because it can't be permuted with any of the genes without changing network. \textbf{b)} Regulon circuit consisting of genes {\it clrA, fiu} and operons {\it entCEBAH, fepA-entD} controlled by two regulators \textit{crp} and \textit{fis} also synchronizes, because symmetry group $S_4$ is conserved irregardless of the number of the regulators. In this case fiber has two regulators and no loops and therefore is characterized by fiber numbers $\rvert n = 0, \ell = 2 \rangle$.
  }
  \label{regulon}
\end{figure}

Beyond operons, the next trivial network structure that can
synchronize genes is the regulon, defined as the set of genes
regulated by a single TF. Figure ~\ref{regulon}a shows an example, the
regulon of the transcription factor Fis in \emph{E.~coli}. The regulon
contains four units: the operon \textit{cbpAM} and the genes
\textit{gltX, gyrB, msrA}.  Gene {\it fis}, is also in turn regulated
by Crp. Crp and Fis are two master regulators involved in a myriad of
functions, the most important is carbon utilization.  This regulon is
an example of a first form of simple symmetry, assuming that the genes
have no other regulators: a simple permutation symmetry of the
regulated genes, for instance \textit{cbpAM} $\leftrightarrow$
\textit{gltx}, or any permutations between the four genes
(Fig.~\ref{regulon}a). This symmetry is described by a symmetry group
called the {symmetric group} $S_n$, which consists of all permutations
of $n$ nodes, in this case $S_4$. This symmetry implies that all the
genes in the regulon are synchronized by trivial coregulation of {\it
  fis}.

When the genes in a regulon are also under the control of other TFs,
then there is also a chance that the regulons would preserve
symmetries.  For instance, the single-regulon circuit controlled by
{\it fis} (Fig.~\ref{regulon}a) can be augmented by a second regulator
as in Fig.~\ref{regulon}b. The same symmetric group $S_4$ describes
the symmetry between the genes {\it clrA, fiu} and operons {\it
  entCEBAH, fepA-entD} since all of them can be permutated. These genes
are  synchronous, but not with the regulators {\it crp}, {\it fis}.

Following the nomenclature for fibers developed in \cite{pnas,ian}, we
characterize these circuits by fiber numbers $\rvert n = 0, \ell
\rangle$ since they have no loops, $n=0$, inside the fiber and $\ell$
external regulators.

It is interesting to compare the circuits found by fibrations with the
most commonly used network motifs.  In the network motif nomenclature
of Alon {\it et al.}, the $\rvert 0,2 \rangle$ fiber shown in
Fig. \ref{regulon}b is called FAN motif \cite{alon}, while the
$\rvert 0,1\rangle$ fiber depicted in Fig. \ref{regulon}a is called a star
motif. We will see next that in order to synchronize the regulator
with its regulon, and extra autoregulation loop is required to induce
a fibration symmetry, leading to the first form of non-trivial fiber
beyond the regulons and simple symmetric groups $S_n$, as we show
next.

\subsection{The autoregulation (AR) loop and regulated genes}

\begin{figure}
  \centering
  \includegraphics[width=\linewidth]{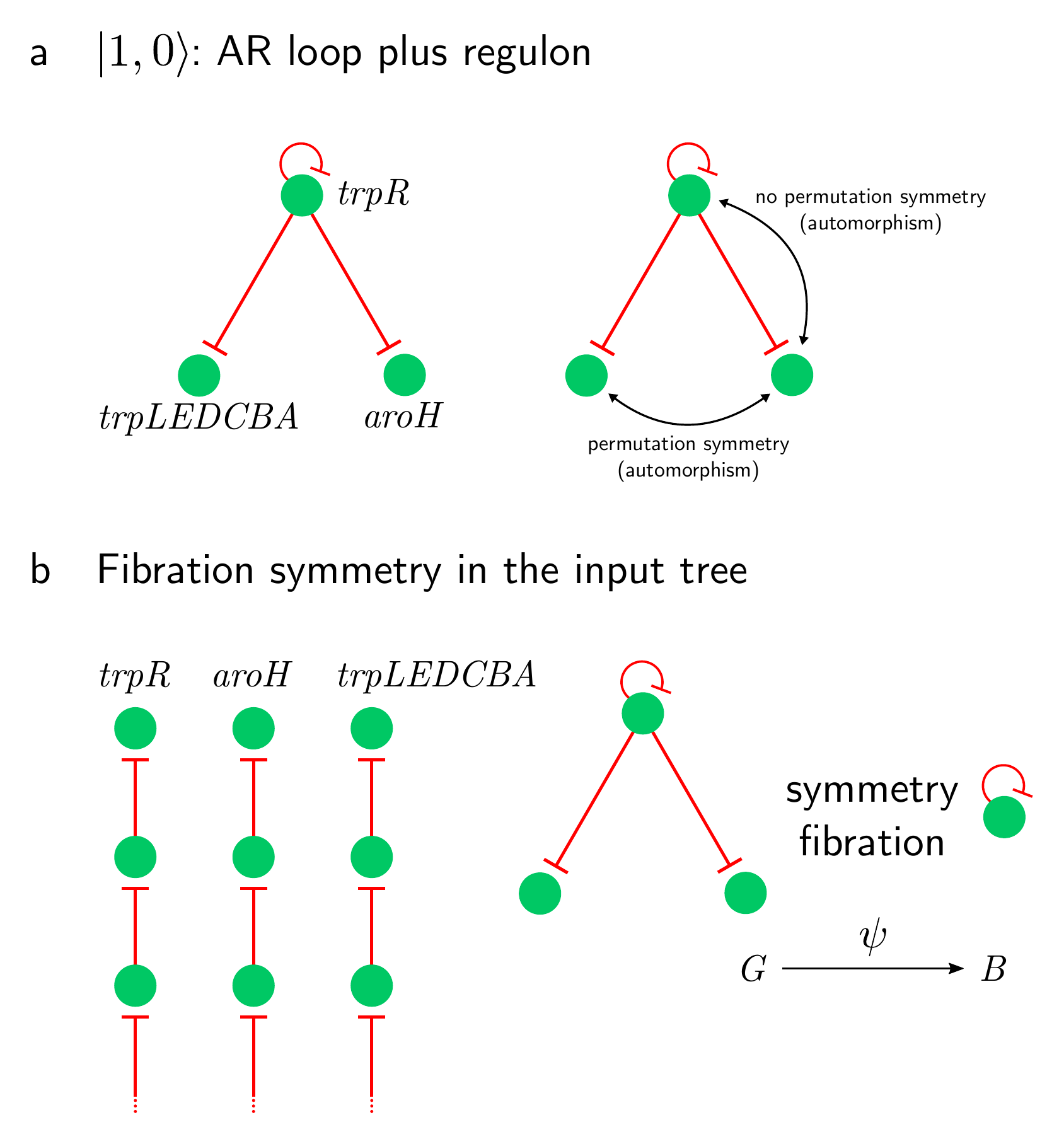}
  \caption{Non-trivial circuits leading to synchronization: AR loop with regulon. \textbf{a)} Genes {\it aroH} and {\it trpLEDCBA} can be permuted under $S_2$ symmetry group, while \textit{trpR} can't be permuted with them without changing the network. \textbf{b)} \textit{trpR} receives input only from itself, therefore its input tree is an infinite chain. {\it aroH} and {\it trpLEDCBA} receive input from \textit{trpR}, that in turn receives input from itself turning these input trees into chains too. Therefore, input trees of all 3 genes are isomorphic to each other. Thus, {\it aroH}, {\it trpLEDCBA} and \textit{trpR} belong to the same fiber and can synchronize their activity. Circuit has one loop and no external regulators, therefore it is classified as $\rvert n = 1, \ell = 0 \rangle$.}
  \label{ar}
\end{figure}

The first non-trivial form of synchronization in the hierarchy of
symmetry fibrations is found when a TF regulates its own expression,
forming an autoregulation (AR) loop, and further regulates other
genes. This is exemplified in Fig.~\ref{ar}a.  In \emph{E.~coli}, such
a circuit is found in the biosynthesis of tryptophan, which is
regulated by TrpR \cite{pmid3106331}, which represses itself, the gene
{\it aroH} (2-Dehydro-3-deoxyphosphoheptonate aldolase), and the {\it
  trpLEDCBA} operon, which codes for the enzymes of the tryptophan
biosynthesis pathway. This circuit is turned on by the presence of
intracellular level of L-tryptophan \cite{Karp:2018aa}. When
tryptophan is in the cell the TF binds and turns off the genes in the
operon.

While the AR loop at {\it trpR} does not affect the automorphisms
formed by the regulated genes (i.e., the regulated genes are invariant
under permutations of the symmetric group S$_2$), the AR loop
introduces the fibration symmetry between {\it trpR}, {\it trpLEDCBA},
and {\it aroH}. This symmetry is not captured by a simple permutation
of the nodes. For instance, permuting the operon {\it trpLEDCBA} with
{\it aroH} preserves the adjacency, but permuting {\it trpR} with
either the operon or {\it aroH} does not preserve adjacency. Therefore
{\it trpR} does not belong to the symmetry group formed by {\it
  trpLEDCBA} and {\it aroH}. Still, as we show next, {\it trpR} is
synchronized with {\it trpLEDCBA} and {\it aroH} by the symmetry
fibration of the input trees. We call this circuit $ \rvert 1,0\rangle$
since it contains an AR loop and no external regulators.

\subsection{The Feed-Forward Fiber}

When an AR circuit such as the {\it trpR}-{\it trpLEDCBA}-{\it aroH}
fiber is regulated by an external TF, it becomes a $\rvert  1, 1 \rangle$
fiber.  This is a prominent circuit in bacteria; we call it the
feed-forward fiber (FFF) \cite{pnas}. The FFF resembles the
feed-forward loop (FFL) network motif introduced in \cite{alon-motif},
except for the additional AR at the intermediate TF TrpR.  This
crucial addition transforms a FFL into a FFF composed of three genes
into a synchronized fiber \cite{ian}.  This type of building block is
abundant in \emph{E.~coli} and \emph{B.~subtilis} GRNs \cite{ian}.

\begin{figure}
  \centering
  \includegraphics[width=\linewidth]{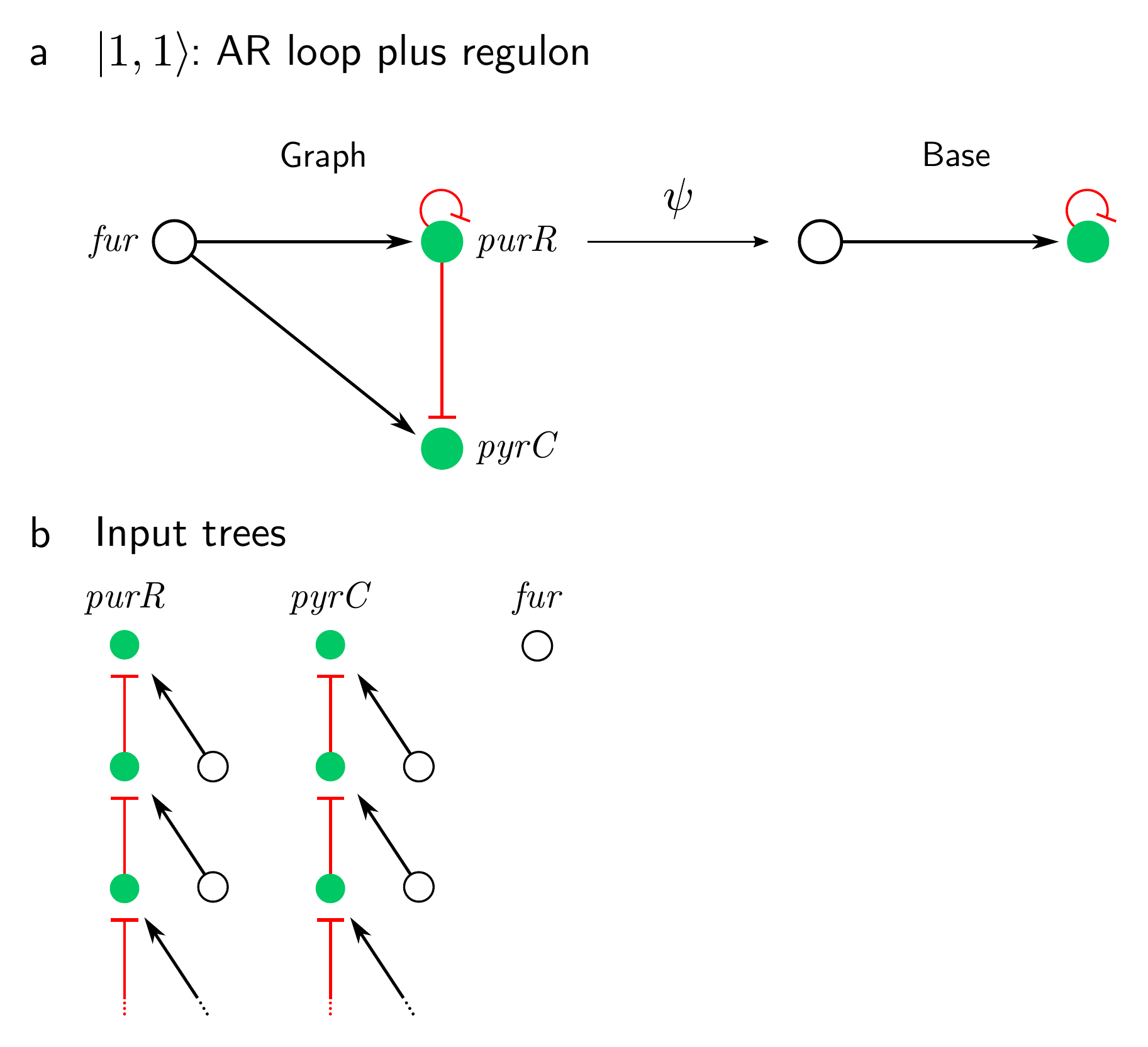}
  \caption{Non-trivial circuits leading to synchronization: FFF (AR loop with regulon and external regulator). \textbf{a)} \textit{purR} and its target gene \textit{pyrC} regulated by {\it fur} form a FFF. FFF has one loop and one external regulator and therefore is classified as $\rvert n = 1, \ell = 1 \rangle$. \textit{purR} and \textit{pyrC} belong to the same fiber (will be shown in \textbf{b)}) and therefore are "collapsed" under fibration $\psi$, while {\it fur} is left untouched. \textbf{b)} \textit{purR} receives an input from itself creating an infinite chain and regulator \textit{fur}, that doesn't have any inputs. Therefore, infinite chain with additional input on each layer represents an input tree of \textit{purR}. Similarly, \textit{pyrC} receives input from \textit{purR} that leads to the infinite chain and \textit{fur} that creates an additional input. \textit{fur} doesn't receive any inputs and therefore has an input tree of height 0. Input trees of \textit{purR} and \textit{pyrC} are isomorphic, therefore \textit{purR} and \textit{pyrC} belong to the same fiber and synchronize their activity.}
  \label{fff}
\end{figure}

An example of an FFF is observed in the purine biosynthesis circuit in
\emph{E.~coli}, shown in in Fig.~\ref{fff}a. It is composed of the
repressor TF \textit{purR} and its target gene \textit{pyrC}, both
regulated by the master regulator {\it fur}.  The input trees of the
genes in this FFF are shown in Fig.~\ref{fff}a. We see that {\it pyrC}
and {\it purR} receive the same inputs from both {\it fur} and {\it
  purR}. On the first layer of the input tree, we find that {\it purR}
is an autoregulator and also regulates the gene.  The second level of
the input tree contains exactly the same genes, and so on. This forms
an input tree of infinitely many layers since there is a loop in the
fiber at {\it purR}.

\subsection{Multilayer composite fiber}

\textcolor{red}{Synchronization of the fiber is defined by the isomorphism between input trees of the nodes in the fiber. Consider the input tree of the FFF building block in Fig.~\ref{fff}. First layer for both $purR$ and $pyrC$ contains the regulator ($fur$) and the green node ($purR$). Therefore, the second layer not only has the same topology, but it has the same nodes. Hence, there is no way to break the isomorphic topology after this layer, because inputs of the same nodes are considered. Therefore, in the case of the FFF building blocks and all the other building blocks considered so far one layer of the input tree or the input set alone is enough to detect synchronization.}
The next level of
complexity in the hierarchy of fibers are circuits where the
synchronization depends on deeper input layers of synchronized genes,
and longer loops of information. This increase in complexity of the
circuits is seen in multilayer composite fibers and Fibonacci fibers
which we treat next.

\begin{figure}
  \centering
  \includegraphics[width=\linewidth]{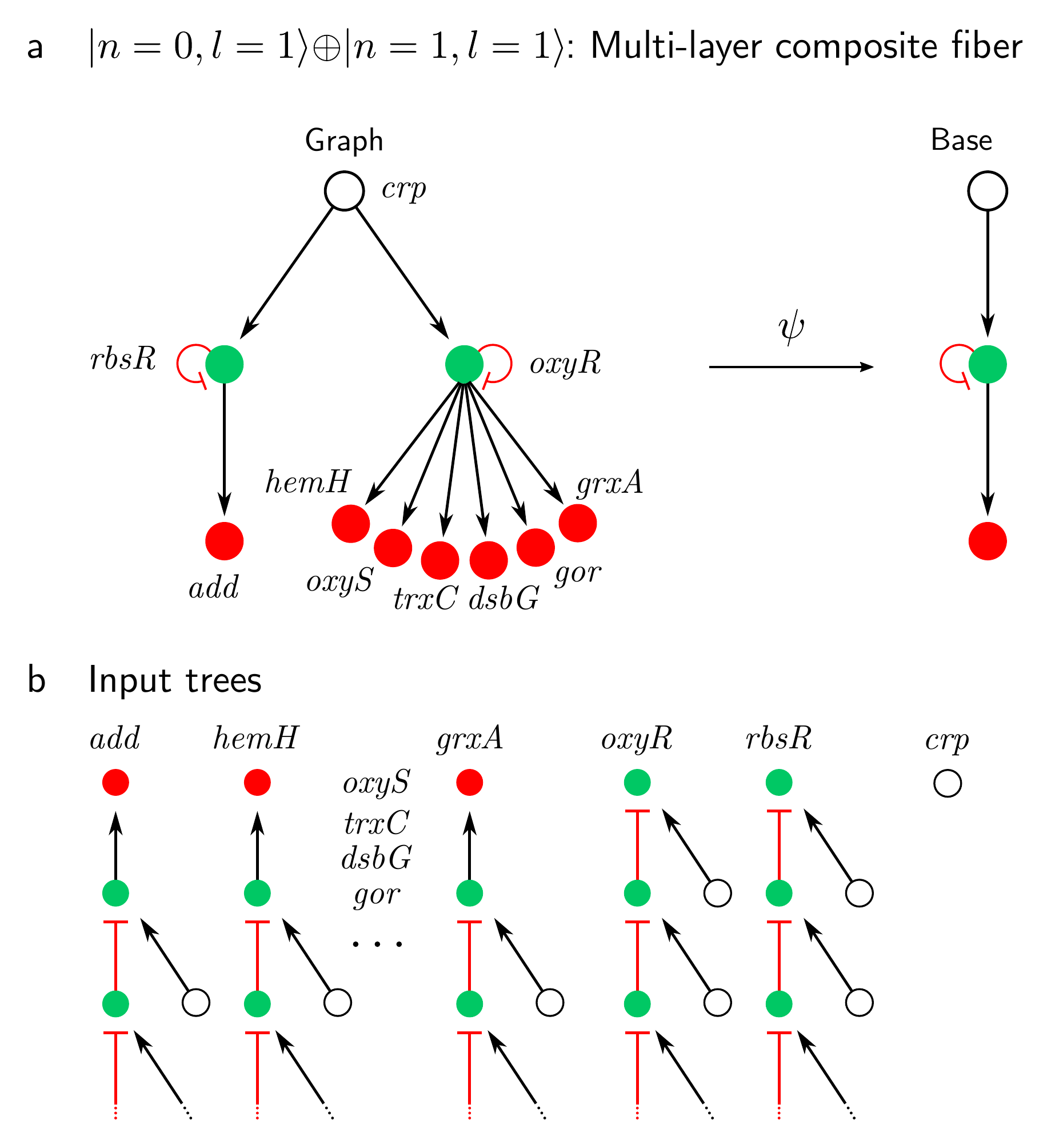}
  \caption{Non-trivial circuits leading to synchronization: Multi-layer composite fiber. \textbf{a)} Circuit consists of two layers of fibers: {\it add, dsbG, gor, grxA, hemH, oxyS, trxC}  classified with $\rvert n = 0, \ell = 1 \rangle$ and {\it rbsR, oxyR} classified with $\rvert n = 1, \ell = 1 \rangle$, therefore forming a multi-layer composite fiber $\rvert n = 0, \ell = 1 \rangle \oplus \rvert n = 1, \ell = 1 \rangle$. Fibration $\psi$ of this circuit "collapses" both fibers and leaves the regulator untouched. \textbf{b)} Genes in the red fiber receive one input from the gene in the green fiber, which in turn receives an input from itself and the regulator. Therefore, input trees of genes in the red fiber resemble the sum of an input tree of $\rvert n = 0, \ell = 1 \rangle$, followed by the input tree of $\rvert n = 1, \ell = 1 \rangle$. Input trees of the green fiber are those of the FFF. Regulator node has no inputs. Thus, multi-layer composite has two non-trivial fibers that can synchronize their activity. Note, gene {\it add} is separated from the rest of the red fiber by two steps, therefore allowing for a long range synchronization in the network.}
  \label{composite}
\end{figure}

Fig.~\ref{composite}a shows an example of a multilayer composite fiber
in \emph{E.~coli} whose main regulator is {\it crp}. In this case,
{\it crp} is the inducer of a composite fiber, composed of {\it oxyR}
and {\it rbsR} and responsible for further downstream regulation of
several carbon utilization subsystems of genes.

The topology of the input trees of this fiber is isomorphic to the one
of FFF $\rvert 1, 1 \rangle$, despite the fact that the building block
has a very different topology than the FFF shown in
Fig.~\ref{fff}a. This first layer of genes regulates via {\it oxyR}
and {\it rbsR} a second fiber composed of genes {\it add, dsbG, gor,
  grxA, hemH, oxyS, trxC}. If the branch corresponding to {\it rbsR}
is disregarded, the building block of the fiber of genes {\it dsbG,
  gor, grxA, hemH, oxyS, trxC} is classified as a single layer $\rvert
0, 1 \rangle$. Thus, the building block corresponding to the entire
fiber in Fig.~\ref{composite}a in red is a double layer composite that
we denote: $\rvert add - oxyS \rangle = \rvert 0, 1 \rangle \oplus
\rvert 1, 1 \rangle$.

Notice that gene {\it add} is two edges apart from the rest of its own
fiber genes, thus achieving synchronization at a distance of two in
the network.


Multilayer fibers are the predominant way for distant synchronization,
indicating the higher level of complexity in these composites
circuits.

\subsection{Fibonacci fibers and the strongly connected component of the network}
\label{sec:ff}

\begin{figure}
  \centering
  \includegraphics[width=\linewidth]{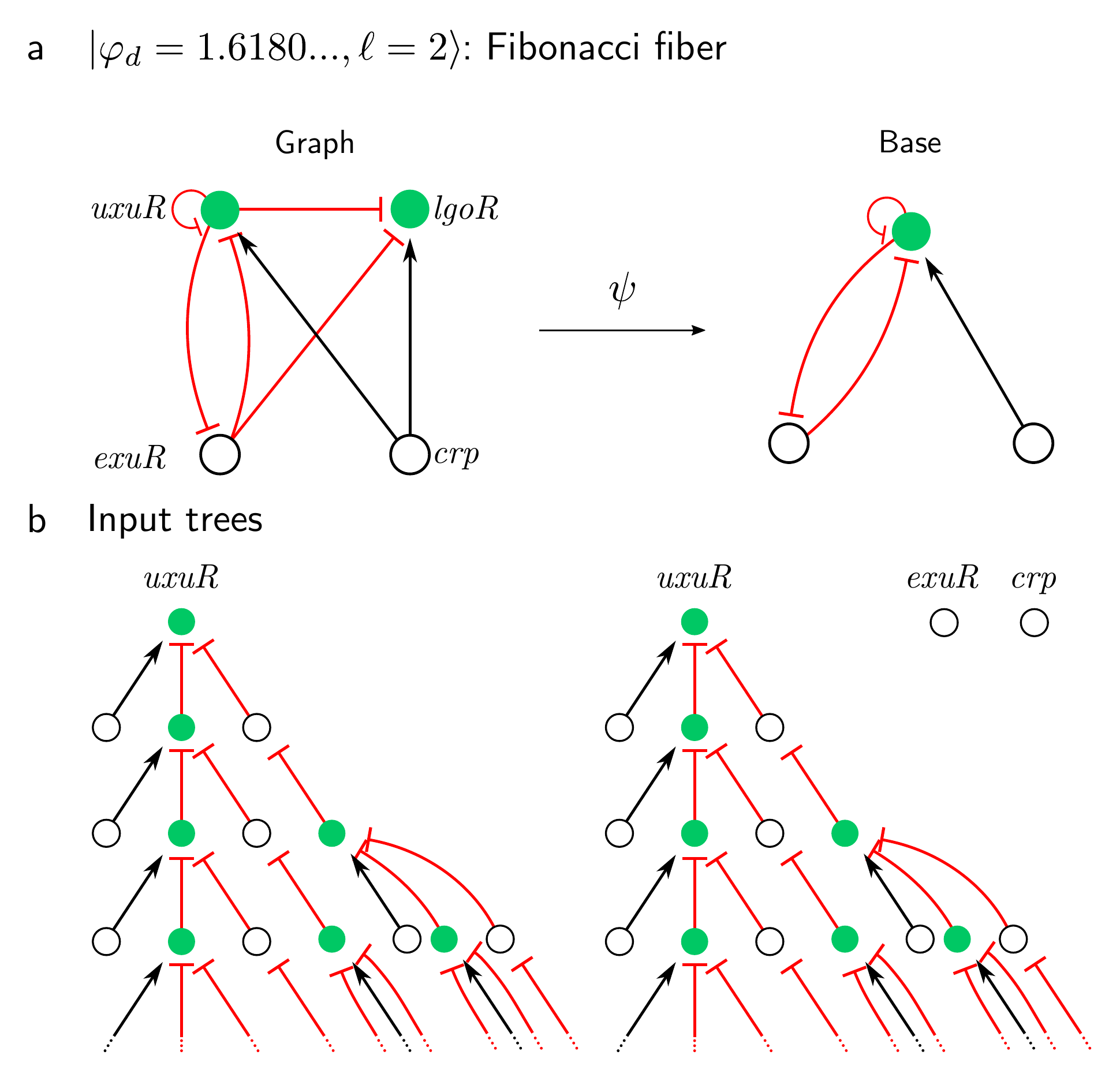}
  \caption{Non-trivial circuits leading to synchronization: Fibonacci
    fiber (FF). \textbf{a)} FF circuit is the FFF circuit with the
    additional edge from the fiber back to the regulator. In this
    example \textit{uxuR} sends back to \textit{exuR}, creating an
    extra loop in the circuit. Extra edge won't change the fiber,
    therefore fibration will stay the same. \textbf{b)} However, extra
    loop changes an input tree of fiber nodes. \textit{uxuR} receives
    from itself and \textit{exuR}, which in turn receives from
    \textit{uxuR}, which creates an input tree with layer sizes
    following Fibonacci sequence. Branching ratio then defines the
    first fiber number and this FF is classified as $\rvert \varphi_d
    = 1.6180..., \ell = 2 \rangle$. Note, node \textit{lgoR} receives
    an input from \textit{exuR} and then from \textit{uxuR}, which
    means that even if there was no link from \textit{uxuR} to
    \textit{lgoR}, information would still be passed along through the
    regulator. This is another way how networks can process the
    information.}
  \label{ff}
  \end{figure}

The next stage in our hierarchy is the Fibonacci fiber (FF) shown in
Fig.~\ref{ff}a \cite{pnas,ian}.  The FF shows a higher level of
complexity in the paths that regulate the fiber. To understand the FF,
we use the concept of strongly connected component, SCC
\cite{pnas}. In general, a fiber may receive information from the
entire network through its input tree. When the fiber is not connected
to a SCC, then information is processed only inside the fiber. This
was the case with all fibers described so far and characterized by
integer fiber numbers $n=0, 1, 2$.

However, if a fiber is connected to a SCC and sends back information
to its own regulators through the SCC, the level of complexity in the
fiber topology increases. In previous examples, the input trees were
infinite due to self-loops, here the input tree becomes infinite due
to longer loops in the SCC of the network, i.e., information cycles
through a longer loop returning back to the fiber. The input tree of
the fiber contains longer loops in the information cycles arriving to
the root gene.  These loops introduce extra terms in the sequence
layers in the input tree leading to Fibonacci sequences in the number
of paths, (see \cite{pnas} for details).  There are an infinite number
of possibilities for these cycles to appear in the network. We find
three types of Fibonacci fibers in the GRN of \emph{E.~coli} in our
previous work \cite{pnas}, one of which is shown in Fig.~\ref{ff}a.
Eukaryotes like yeast and humans present a much richer variety of
Fibonaccis as shown in \cite{ian}.


\subsection{$\rvert n=2, \ell \rangle$: binary tree fiber. }

The last type of building block in the hierarchy of fibers is
characterized by two AR loops, leading to a symmetric input tree (such
a circuit is shown in Fig.~\ref{hierarchy}).  The sequence of
information is coded in a sequence defined by $a_i = 2 a_{i-1}$ and we
classify this fiber as $n=2$.  This procedure can be iterated to any
number of loops, but in the studied bacterial networks we did not find
any fiber with $n>2$, suggesting a practical limit in complexity in
these organisms.

\section{Synchronized coexpression within gene fibers -- experimental validation}
\label{results}

We saw that gene fibrations, in theory, can lead to synchronization.
To see whether this prediction takes place in reality, we now consider
the gene fibers uncovered in bacteria and confront the predicted
coexpression structures with experimental transcriptome data (for
details see Methods section \ref{sec:MethodsDataSets}). We use the
gene expression compilations from Ecomics \cite{ecomics} (for
\textit{E. coli}) and SubtiWiki \cite{SubtiWiki} (for
\emph{B.~subtilis}).  The Ecomics portal collects microarray and
RNA-seq experiments from different strain and sources including NCBI
Gene Expression Omnibus (GEO) public database~\cite{geo} and
ArrayExpress~\cite{array}.  The data is also compiled at the Colombos
web portal~\cite{colombos}.  We choose Ecomics over Colombos because
Ecomics provides absolute expression levels.  Datasets for gene
expression like Colombos \cite{colombos} do not provide the absolute
expression levels but the fold change from a wild-type to a
perturbation state such as a mutation. Measuring the fold change does
not allow to test the synchronization in fibers since the prediction
of the theory refers to unperturbed states in the wild type.  Thus, we
base our analysis on the wild-type networks, expression data from
mutant strains were not taken into account, since mutations leads to
breaking of symmetries.  Theoretical predictions for the response to
mutations will be studied elsewhere.

Subtiwiki is a comprehensive knowledge base for bacterium {\it
  B. subtilis} containing expression, pathways, interactions and
regulation data in the wild-type strain across different experimental
conditions.  A test for gene synchronization in fibers has been
performed in our previous study in \cite{pnas} by looking at specific
experimental conditions where the genes have been activated. Below, we
test the existence of fibers in a larger context with and without
activated conditions and test
the statistical significance of these correlations, as assessed by
p-values.

We study gene expression data using Pearson coefficient of correlation. To find Pearson coefficient of correlation
between gene expression profiles of genes $i$ and $j$ we use:

\begin{equation}
C(i,j)= \dfrac{1}{T} \sum_{t=1}^{T}
\Big(\frac{x_{i,t}-\mu_i}{\sigma_i}\Big)\Big(\frac{x_{j,t}-\mu_j}{\sigma_j}\Big),
\end{equation}

where $T$ is the number of conditions, $x_{i,t}$ is the expression value of gene
$i$ for condition $t$, and $\mu_i$ and $\sigma_i$ are the respective
mean and standard deviation of expression values of gene $i$ for all
conditions.

\begin{figure*}
  \centering
  \includegraphics[width=0.8\linewidth]{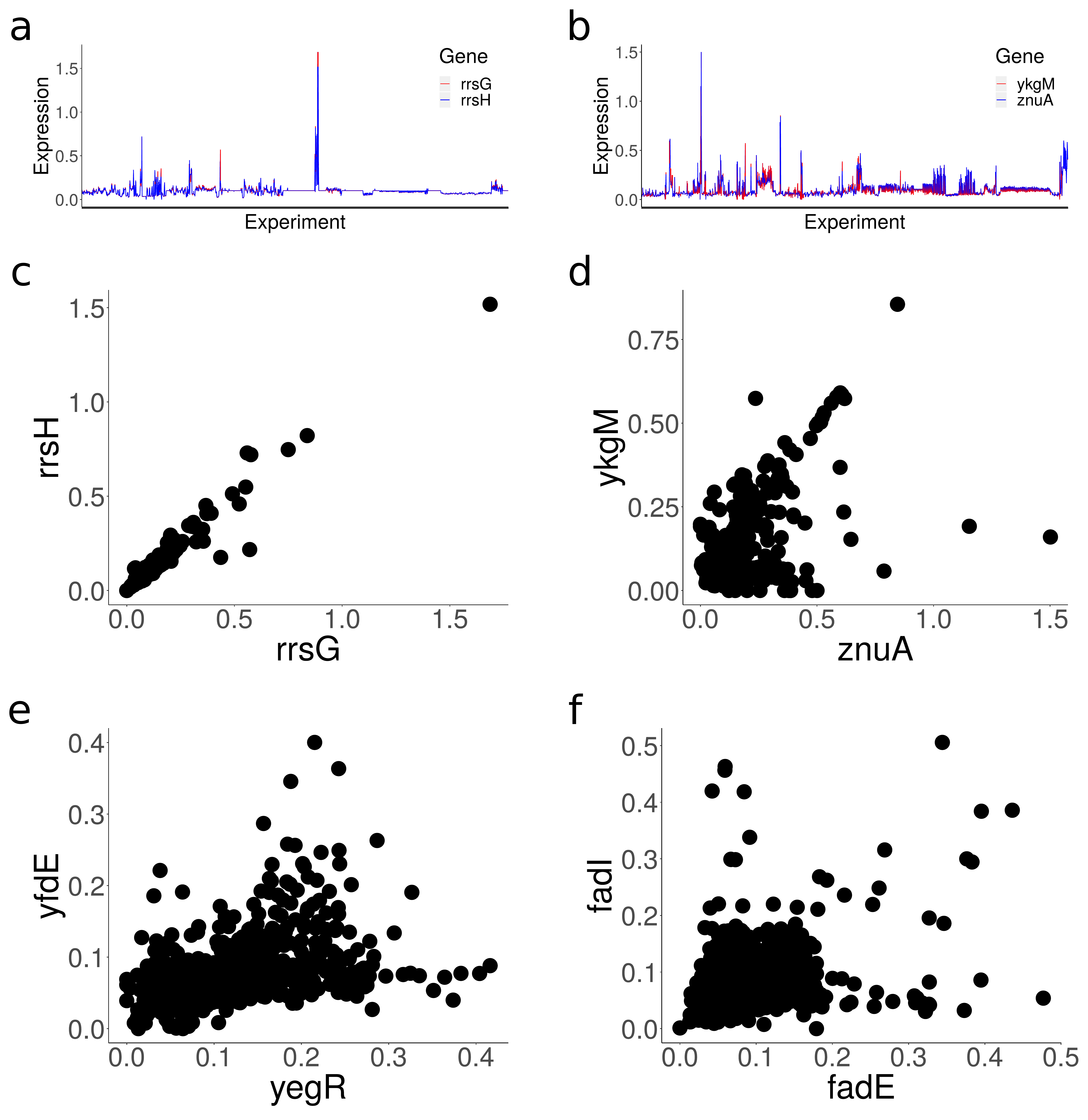}
  \caption{Similarity in gene expression data for selected pairs of genes belonging to the same fiber.
  Gene co-expression is demonstrated on the data from all experiments in the Ecomics database.
  We pick best examples out of 85 fibers obtained in {\it E. coli}.
  \textbf{(a, b)} Gene expression of pairs of genes in {\it rrsH, rrsG} and {\it ykgM, znuA} for 1575 experimental conditions from Ecomics.
  It's easy to see that data is highly correlated.
  %
    \textbf{(c)} Gene expression of {\it rrsH} (Position in genome: 223771--$>$ 225312)
    vs gene expression of {\it rrsG} (Position in genome: 2729616 $<$-- 2731157), correlation =
    0.98,
    \textbf{(d)} Gene expression of {\it ykgM} (312514 $<$-- 312777)
    vs gene expression of {\it znuA} (1941651 $<$-- 1942583), correlation = 0.58,
    \textbf{(e)} Gene expression of {\it yfdE} (2488023 $<$-- 2489168)
    vs gene expression of {\it yegR} (2167989$<$-- 2168306), correlation = 0.49,
    \textbf{(f)} Gene expression of {\it fadI} (240859 $<$-- 243303)
    vs gene expression of {\it fadE} (2459159 $<$-- 2460469), correlation = 0.49.}
  \label{Fig:Significance1}
\end{figure*}

We start by considering expression of few pairs of nodes that are
predicted to be synchronized by fibration theory in {\it
  E. coli}. Figure \ref{Fig:Significance1} shows the gene expression
correlations between four different gene pairs, each from one fiber,
in the form of scatter plots. Each plot can be quantified by a single
Pearson correlation value. Fig.~\ref{Fig:Significance1}c depicts the
expression of \textit{rrsH} vs \textit{rrsG} which form a fiber, and
are predicted to be synchronized. Each point in the plot represents a
single experiments with a particular growth condition for the
bacterium as obtained from the Ecomics dataset.  The observed Pearson
correlation value is 0.98 which indicates a strong synchronization in
the activity of the genes across the experimental conditions. These
genes are located far away in the genome and as it can be seen from
the scatter plot their expression is highly
correlated. Fig.~\ref{Fig:Significance1}d-f shows another few scatter
plots of gene expression in {\it E.coli}. For instance,
fig. \ref{Fig:Significance1}f shows the correlation between {\it fadI}
and {\it fadE} which also form a fiber with a correlation coefficient
0.49.  These examples are highly correlated, but as it can be seen
from Fig.~\ref{Fig:Significance1} there is a lot of noise.

In the following sections we assess whether the observed correlations
are significantly large within the predicted fibers  by assessing
gene correlations in the entire data set, within and between gene
fibers. We report two kinds of correlations: (a) correlations computed
from the entire data set, that is, using all the experimental
conditions appearing in Ecomic for {\it E. coli} without filtering
irrespective of whether the genes are being expressed in the
particular conditions or not, and (b) correlations obtained after a
filtering of ``active experimental conditions'', specifically chosen
for each set of genes in the fibers.  This second approach has been
used in our previous study in \cite{pnas} and it is similar to the
filtering method employed by the Colombos database at \cite{colombos}.
Then, we scrutinize the statement that fibrations predict larger mean
correlation (i.e. gain of synchronization) in the fibers of {\it
  E.coli}.  Again, we first give significant results without filtering
and we continue by applying the variation of the filtering method used
by Colombos \cite{colombos} and showing significance and results of
that method.

\subsection{Correlations within and between gene fibers}

In this subsection we give an overview over the correlations between
all genes in the dataset, and their relation to gene fibers. We start
by grouping the genes in the predicted fibers (85 in total in {\it
  E. coli}) and then using all the experimental conditions in Ecomics
(1575 conditions) to calculate the correlation matrix between and
within genes in fibers to test for synchronization. Later we will
filter these conditions to those where the genes are active.

\begin{figure*}[t!]
  \centering
  \includegraphics[width=\linewidth]{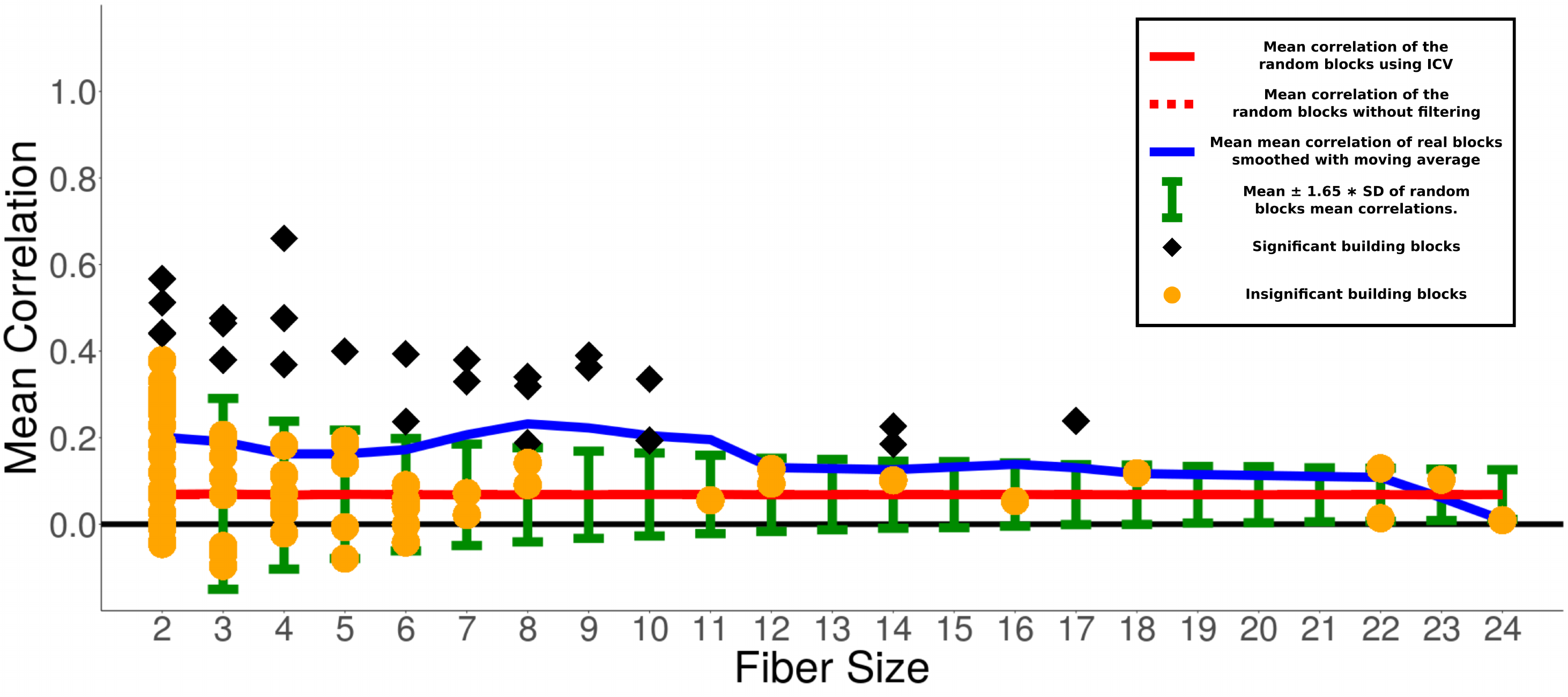}
  \caption{Mean correlation within the fibers, computed without filtering versus sizes of gene fibers (number of nodes). Black and orange dots - mean correlation of 85 fibers of size $<$
    25. Shape shows significance: black diamond - significant, orange circle - insignificant. Blue - mean of mean
    correlation of real blocks in black (smoothed with moving
    average). Green error bars - $mean \pm 1.65 * SD$ of random blocks
    mean correlations. $1.65$ is chosen, because $0.05$ values of the normal distribution with $\mu=0$ and $\sigma=1$ are above $1.65$, therefore $1.65$ corresponds to the p-value of 0.05.
    Red - mean correlation of the random blocks.}
  \label{Fig:Significance2}
\end{figure*}

In order to quantify the synchronization in the fibers we consider the
statistics of mean correlations inside the fiber, where mean
correlation is the mean of all off-diagonal terms in the correlation
matrix. To assess the statistical significance of the correlations we
compare the fiber mean correlations with mean correlations of random
sets of nodes of the same size. We assume that mean correlations of
random blocks of a given size are distributed normally. We define mean
and standard deviation of the random set by finding mean correlation
of 100,000 random sets of fibers of size ranging from 2 to 24 genes
(as found in {\it E. coli}) using all the conditions of expression
data. Then we find mean and standard deviation of this 100,000 random
sample. Summary of this analysis is shown in
Fig.~\ref{Fig:Significance2}. In general we find that the mean
correlation for the genes inside the fiber is relatively low, with all
fibers with mean correlation below 0.5.
We will see below that this is due to the fact that in many experimental conditions from Ecomic the fibers are not activated.
However, it is also clear from the data that
there is an increase in correlation inside the fibers (blue curve in
Fig.~\ref{Fig:Significance2}) as compared with the mean correlation
in the random sets (red curve in Fig.~\ref{Fig:Significance2}), but
how significant is this increase?

In order to quantify the statistical significance of the increase in
correlations in fibers, we study the probability that random sets of
genes of size $n$ have the mean that is higher or equal than the mean
correlation of fibers of the same size $n$. We then calculate the p-value
of the measured distribution belonging to the random distribution.

The sampling distribution of a normal distribution of size $m$ has the
mean distributed normally with mean ($\mu_m=\mu$) and standard
deviation $\sigma_m = \frac{\sigma}{\sqrt{m}}$, where $\mu$ and
$\sigma$ are mean and standard deviation of the original distribution
\cite{cramerStatistics}. For a normal distribution with known mean and
standard deviation, we find a z-score that corresponds to our
measurement, i.e. to the mean of mean correlations of fibers
$\mu_{real}$ as $Z = \frac{\mu_{real} - \mu_m}{\sigma_m}$ and the
corresponding p-value. Table \ref{Table:SignificanceNoFilter} shows
the summary of these measurements. We find that 68 out of 85 fibers
are significant, which indicates that for most of the fibers the mean
correlation is significantly higher than random. Therefore fibers
significantly predict the gain of synchronization even in this
unfiltered dataset. However the typical mean correlation is low enough
(below 0.5) to consider this result of significance.

\begin{table*}[ht]
  \centering
  \begin{adjustbox}{width=0.8\textwidth}
    \begin{tabular}{|C|L|L|M|C|C|}
      \hline
      Fiber size & Mean of mean correlations ($\mu_{real}$) & Mean of random mean correlations ($\mu_m$) & Standard deviation of random mean correlations ($\sigma_m$) & Number of blocks ($m$) & p-value \\ 
      \hline
      2 & 0.20 & 0.07 & 0.20 & 24 & \textbf{0} \\ 
      3 & 0.17 & 0.07 & 0.13 & 11 & \textbf{0.01} \\ 
      4 & 0.20 & 0.07 & 0.10 & 10 & \textbf{0} \\ 
      5 & 0.14 & 0.07 & 0.09 & 6 & \textbf{0.03} \\ 
      6 & 0.10 & 0.07 & 0.08 & 8 & 0.11 \\ 
      7 & 0.20 & 0.07 & 0.07 & 4 & \textbf{0} \\ 
      8 & 0.22 & 0.07 & 0.07 & 5 & \textbf{0} \\ 
      9 & 0.38 & 0.07 & 0.06 & 2 & \textbf{0} \\ 
      10 & 0.26 & 0.07 & 0.06 & 2 & \textbf{0} \\ 
      11 & 0.06 & 0.07 & 0.05 & 1 & 0.60 \\ 
      12 & 0.11 & 0.07 & 0.05 & 2 & 0.12 \\ 
      14 & 0.17 & 0.07 & 0.05 & 3 & \textbf{0} \\ 
      16 & 0.05 & 0.07 & 0.04 & 1 & 0.63 \\ 
      17 & 0.24 & 0.07 & 0.04 & 1 & \textbf{0} \\ 
      18 & 0.12 & 0.07 & 0.04 & 1 & 0.11 \\ 
      22 & 0.07 & 0.07 & 0.04 & 2 & 0.45 \\ 
      23 & 0.10 & 0.07 & 0.04 & 1 & 0.17 \\ 
      24 & 0.01 & 0.07 & 0.03 & 1 & 0.95 \\ 
       \hline
    \end{tabular}
  \end{adjustbox}
  \caption{Significance of the increase in correlation obtained using method with no filtering. P-values $<$ 0.05 in bold. 68/85 are significant. Random sample consists of 100000 fibers.}
  \label{Table:SignificanceNoFilter}
\end{table*}

\subsection{Inverse Coefficient of Variation to filter out conditions of gene expression}
\label{ICV}

\begin{table*}[ht]
  \centering
  \begin{adjustbox}{width=0.8\textwidth}
    \begin{tabular}{|C|L|L|M|C|C|}
      \hline
      Fiber size & Mean of mean correlations ($\mu_{real}$) & Mean of random mean correlations ($\mu_m$) & Standard deviation of random mean correlations ($\sigma_m$) & Number of blocks ($m$) & p-value \\ 
      \hline
      2 & 1 & 0.99 & 0.08 & 24 & 0.32 \\ 
      3 & 0.81 & 0.78 & 0.17 & 11 & 0.25 \\ 
      4 & 0.78 & 0.58 & 0.22 & 10 & \textbf{0} \\ 
      5 & 0.53 & 0.46 & 0.22 & 6 & 0.24 \\ 
      6 & 0.44 & 0.38 & 0.20 & 8 & 0.20 \\ 
      7 & 0.50 & 0.32 & 0.19 & 4 & \textbf{0.03} \\ 
      8 & 0.51 & 0.28 & 0.18 & 5 & \textbf{0} \\ 
      9 & 0.77 & 0.25 & 0.17 & 2 & \textbf{0} \\ 
      10 & 0.55 & 0.22 & 0.16 & 2 & \textbf{0} \\ 
      11 & 0.25 & 0.20 & 0.14 & 1 & 0.37 \\ 
      12 & 0.20 & 0.18 & 0.14 & 2 & 0.44 \\ 
      14 & 0.28 & 0.15 & 0.12 & 3 & \textbf{0.03} \\ 
      16 & 0.07 & 0.14 & 0.11 & 1 & 0.74 \\ 
      17 & 0.50 & 0.12 & 0.10 & 1 & \textbf{0} \\ 
      18 & 0.19 & 0.11 & 0.09 & 1 & 0.21 \\ 
      22 & 0.10 & 0.10 & 0.08 & 2 & 0.43 \\ 
      23 & 0.23 & 0.09 & 0.07 & 1 & \textbf{0.03} \\ 
      24 & 0.02 & 0.09 & 0.07 & 1 & 0.84 \\ 
      \hline
    \end{tabular}
  \end{adjustbox}
  \caption{Significance of the increase in correlation obtained using
    ICV. P-values $<$ 0.05 in bold. 28/85 are significant. Random
    sample consists of 100000 fibers.}
  \label{Table:SignificanceICV}
\end{table*}

\begin{figure*}[t!]
  \centering
  \includegraphics[width=\linewidth]{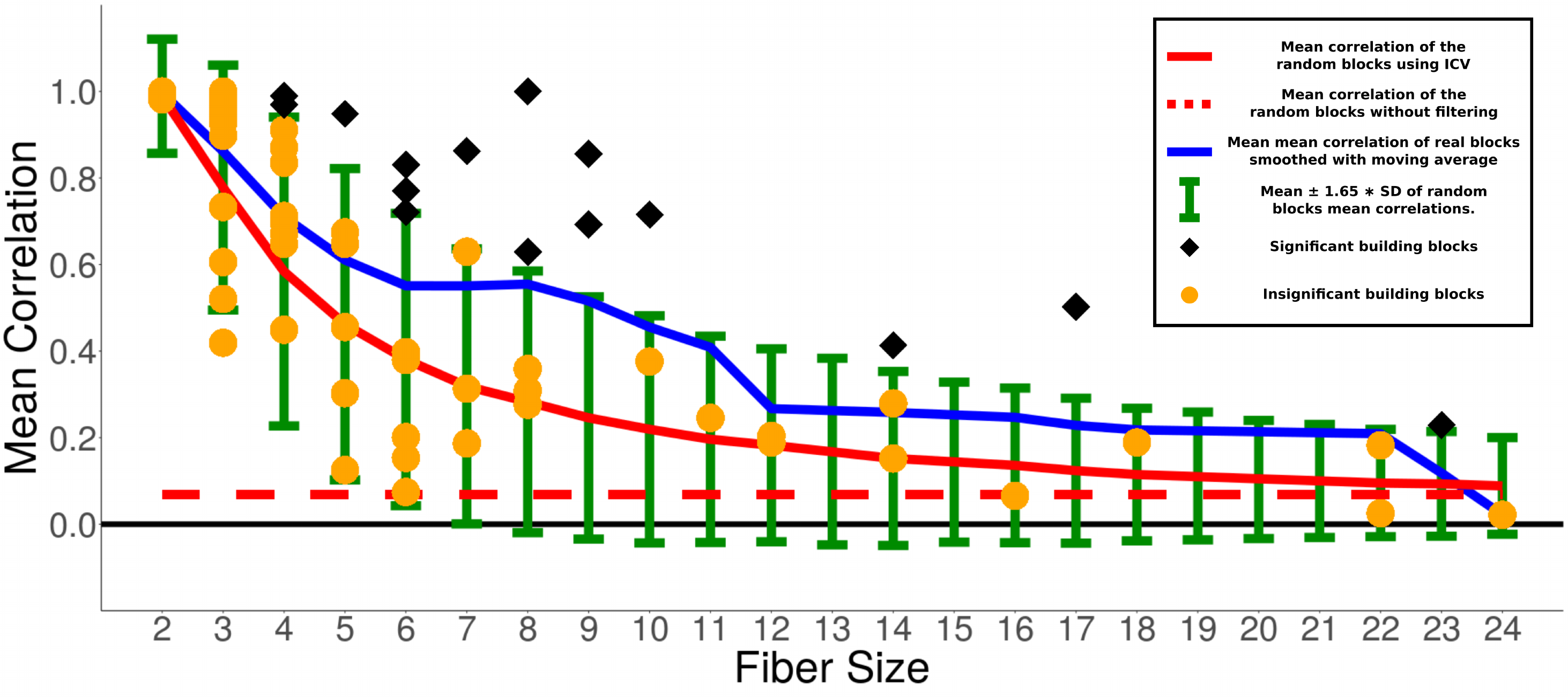}
  \caption{Mean correlation of blocks calculated using the method of ICV vs number of nodes in the fiber.
  Black and orange dots - mean correlation of 85 fibers of size $<$
    25. Shape shows significance: black diamond - significant, orange circle - insignificant. Blue - mean of mean correlation of real blocks in black
    (smoothed with moving average). Green error bars - $mean \pm 1.65 *
    SD$ of random blocks mean correlations. $1.65$ corresponds to the p-value of 0.05 as explained in Fig.~\ref{Fig:Significance2}. Red solid line - mean
    correlation of random blocks found using ICV. Red dashed line -
    mean correlation of the random blocks without filtering.}
  \label{Fig:Significance3}
\end{figure*}

The results so far refer to gene correlations across the entire data
set, including all experimental samples. However, in our theoretical
predictions we assume that correlations in a fiber should only exist
when the fiber is ``active'', that is, in a subset of ``active''
experimental conditions, which is specific to this fiber. Since we
don't have any other information about these active conditions, next,
we determine a set of active conditions, for each single fiber, by a
heuristic criterion, based on the ICV (Inverse Coefficient of
Variation, see Methods). It is important to note that this filtering
for high ICV values is, basically, also a filtering for conditions in
which the genes tend to be correlated: i.e.~our filtering for active
conditions will induce correlations. To see which of these
correlations represent ``real'' correlations and not just statistical
artifacts, we rely again on our permutation test.

Indeed, cells can adapt to varying growth conditions by sensing
extracellular cues, using intracellular effector molecules to modulate
transcription factor activities. This activation can happen directly
or via intermediate signaling pathways such as the two-component
systems (TCS) in bacteria. Therefore, different genes will not be
expressed all the time, but will only be expressed under specific
external experimental growth conditions. We consider that a gene is
either active (expression much larger than noise range) or inactive
(expression in the noise range).  Furthermore, the activity of all TFs
are modulated by effectors (ligands and metabolites). In our approach,
these additional regulations are ignored and requiere a detailed
consideration of the metabolic networks that is coupled to the
GRN. Such coupling will be studied in a forthcoming paper.  In the
present analysis we consider that these effectors activate and
deactivate the fiber circuits identified by fibrations and are
determined by the internal metabolism of the cell and the external
growth conditions where fibers are activated.
  
A specific example of activation mediated through a known effector is
the cAMP activation of {\it crp}. When the genes are not significantly
expressed or expressed below the noise level, the corresponding
activity in expression is expected to be random noise.  When the genes
are active or significantly expressed for a given statistical test of
significance, the genes can be coexpressed by showing large
correlations in their expression levels or they can not be
coexpressed, by showing zero or near noise level correlations in their
expression levels over time.

When the expression correlations between genes are computed from the
entire set of conditions as done in the previous section, the noise in
the conditions where genes are inactivated distorts the
results. That is, using the inactivated states to compute the
correlations, there will be noise, which makes it hard to detect
correlations, leading to the need to filter the conditions.

Thus, to test coexpression patterns in a given set of genes, we first
find the conditions under which this set of genes is active, ie,
over-expressed under a given statistical test.  Then we assess, just
for these conditions, the coexpression between our genes.  Being
inactivated together by external effectors, we filter out for
experimental growth condition where the fibers are activated and
present correlations over a given determined threshold of noise. We
note that activation of genes does not imply correlations per
se. Thus, there are two different stages in the analysis that are
subject to different statistical tests of their significance as we
explain below.

For this study we use Ecomics since it provides the data in the
wild-type (WT) conditions, rather by providing the data using the
fold-change that compares a mutation or perturbation to the WT. Using
Ecomics \cite{ecomics}, we obtain the set of experimental conditions
where the particular genes in a given fiber have been significantly
expressed. For this task we follow standard gene expression analysis similar to the one
developed in \url{colombos.net} and Ref. \cite{colombos} for the
expression levels in \emph{E. coli} to first identify the set of
growth condition where the genes in a given fiber are significantly
expressed respect to random noise and then test the synchronization
through correlations in gene's activity using these conditions.  We
then repeat the scheme using the conditions in Subtiwiki for {\it
  B. subtilis} \cite{SubtiWiki}.

To choose our gene sets to test for synchronization, we first consider
the building block of the fiber. We test the synchronization in the
fiber of the building block and the lack of synchronization between
the fiber and its external regulators. We then consider the cross-correlations between fibers.

For a given set of genes in a fiber, we find the experimental
conditions for which the genes have been significantly expressed by
comparing the expression samples over different biological
conditions. To filter conditions where the genes are expressed we use
the Inverse Coefficient of Variation (ICV) similar to the one applied
by Colombos \cite{colombos}. We consider the genes in the fiber and
obtain their expression levels for all conditions. Then we calculate
the ICV for all conditions using the method explained in Section
\ref{sec:MethodsICV}.

After selecting the conditions for expression according to the
relevant ICV, we use the expression data for the selected experimental
conditions and we find the Pearson correlation coefficient of
correlation between gene expression profiles of genes $i$ and $j$. For
genes that are in the same fiber, we calculate the correlation matrix
averaging over the experimental conditions of the fiber using the ICV
method explained above.  To compute correlations between  genes
 belonging to different fibers, we consider the correlation function
calculated over the union of conditions used for two fibers.

The above framework yields a correlation matrix for a fiber and its
regulators. We apply the method to each fiber in the networks of {\it
  E. coli} and {\it B. subtilis} to test the prediction that genes in
the fiber are more correlated with genes in the fiber, than with genes
outside the fiber.

To deal with the noise generated by the inactivated states of the
genes, we filter the conditions based on the ICV \cite{colombos}.  ICV
allows us to consider only conditions where mean expression is few
standard deviations higher than 0, that is $\mu_{expression} > n *
\sigma_{expression}$, where n is an arbitrary number (see Methods). In
other words, we consider conditions where the fiber is activated. In
principle, this could create a bias towards increased correlation, so
the question arises of how significant results obtained using this
method are.

The analysis of gene correlations with ICV filtering was performed as
follows.  We considered each fiber, determined the active conditions
for this fiber, and computed the intra-fiber correlations over this
set of conditions. To compute gene correlations across two different
fibers, we considered the previously determined active conditions for
both fibers and computed the correlations over these conditions.

Figure~\ref{Fig:Significance3} shows the summary of the
results similar to the one presented before for the method with no
filtering by comparing the significance of the correlations in the
fibers with a null model of 100,000 random set of genes with sizes
from 2 to 24. First, we observe that mean correlation in the random
set approached 1 when the size of the set approaches 2. This implies
that for fibers with two genes, the filtering method is not
significance, i.e., any random set of genes will show high
correlations after filtering the conditions with ICV. However, the
average of the within-fiber mean correlation for a given size of fiber
as a function of the size of the fiber (red curve in
Fig.~\ref{Fig:Significance3}) slowly decays towards the red dashed
line, which means that the correlation bias created by ICV disappears
for bigger sizes of the fiber. Second, we can see a clear increase of
mean correlation (blue line being higher than red) similar to the one
we observed in the method with no filtering. This implies that we
again see the increase in correlation within the fiber using the
method of ICV. Significance of this increase can be studied observing
Table \ref{Table:SignificanceICV}. We see that 28 out of 85 fibers are
significant. The proportion of the blocks that are significant using ICV
method is lower than without filtering, which is mostly happening due
to the fact that the increase in the correlation for the smaller sizes
is less significant.

Having studied the correlations across all genes in fibers with
filtering and no filtering of conditions, we conclude that there is a
clear pattern of increase of correlation between genes in the fiber.
Method with no filtering shows that small fibers detect the
significant gain of synchronization, while method with filtering shows
the gain of synchronization in bigger fibers. It's important to note
that there are fibers that are significant and highly correlated
individually as it can be seen from Fig.~\ref{Fig:Significance2} and
Fig.~\ref{Fig:Significance3}, but what we observe here is a
consolidative effect of gain of synchronization, although with extra
fiber-fiber correlations which are not directly related to the fiber
structure but may indicate activation of different fibers under same
conditions.  We shall now describe some particular cases of fibers
with biological functionalities.





\begin{figure*}
  \centering
  \includegraphics[width=0.85\linewidth]{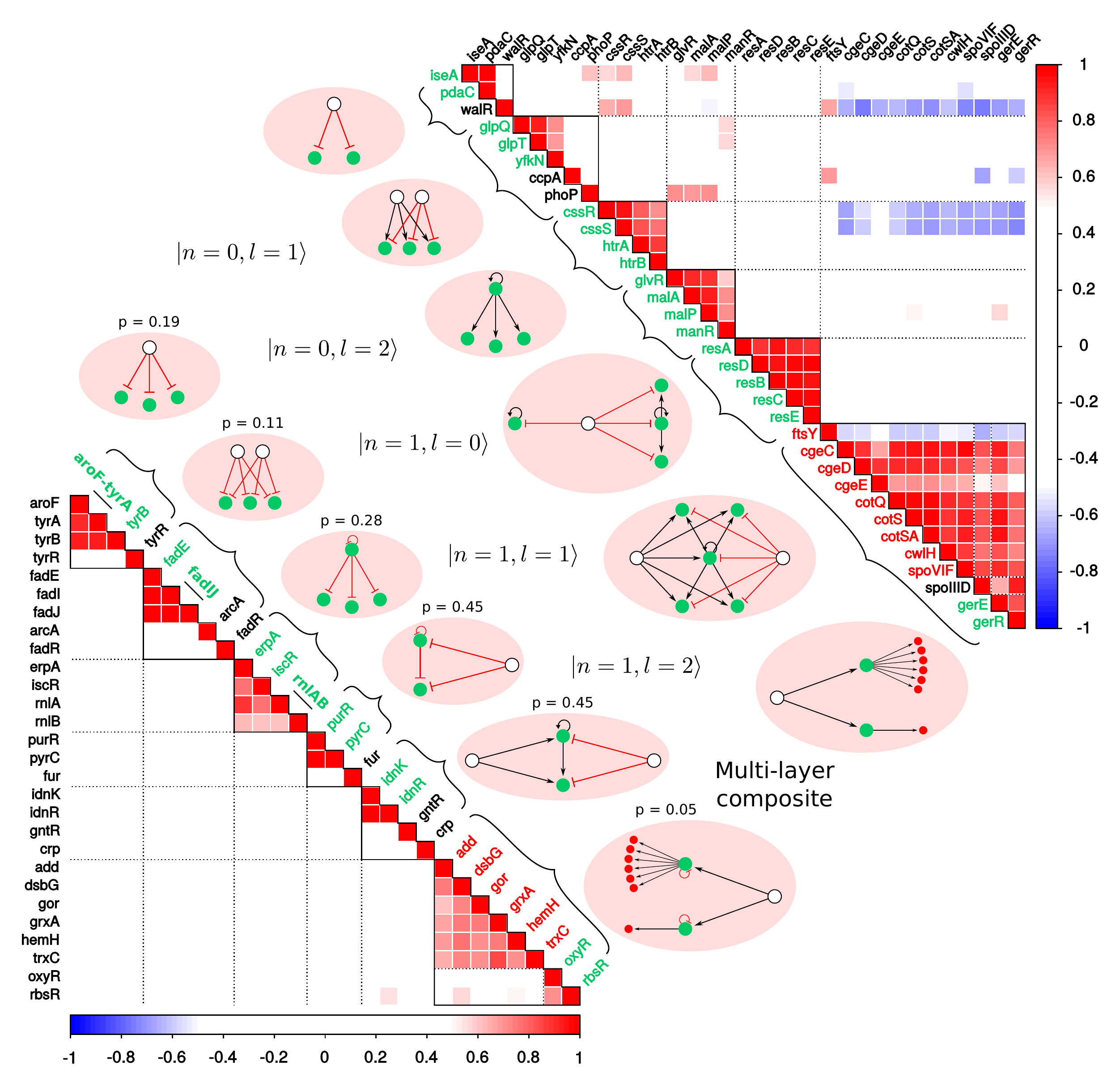}
  \caption{Fiber Class Examples. We display the six
    different fiber classes with their genetic circuit and
    correlation matrix.
    \textbf{Genetic Circuits:} A
    graphical representation of the genes and their regulators
    interactions. Edges: Black - Activation, Red - Repression. Nodes: Green and red - Fibers, White - Regulators.
    \textbf{Correlation Graphs:} Correlation between Fiber genes (green and red font)
    and regulators (black font). Operons are shown with lines along the correlation matrix diagonal. Black lines enclose fibers, black dotted lines show cross-correlations between fibers and inside multi-layered fibers. Genes inside fibers are correlated and are not correlated with regulators and different fibers. Note, observed correlations have high p-values using ICV method. This happens due to the fact that displayed fibers are small and as mentioned before small fibers are have high p-values with method with no filtering.
    Observed correlations can guide future research in finding missing transcriptional regulations. For example, self-regulation loop on \textit{spoIIID} could explain the correlation inside multi-layered fiber in bacillus.}
\label{hierarchy}
\end{figure*}

\subsection{Fiber synchronization in the hierarchy of fibers in E. coli and B. subtilis at the network level}

So far we have tested coexpression within fibers separately,
selecting for each fiber the conditions under which the genes in this
fiber are activated. \textcolor{red}{We then consider cross-correlation between fibers to check the validity of our results.}
A synchronization across fibers might
indicate that our gene fibers are not correct, maybe because of
missing edges in the reconstructed network.

To calculate the correlations inside
fibers we use the conditions where each fiber has been overexpressed
as given by the ICV. In general, these sets of conditions need not
overlap for two different fibers. For this reason, when calculating
the off-diagonal correlation between different fibers we consider the
union of the conditions of both fibers to calculate the correlation
matrix.

Figure \ref{hierarchy} shows the expression correlation matrix for
genes in a number of circuits in \emph{E.~coli} and
\emph{B.~subtilis}, following the hierarchy of fibers explained in
Section \ref{sec:hierarchy}.  We study the following fibers: $\rvert
0, \ell \rangle$ with $\ell=1, 2$, followed by three examples of FFF
$\rvert 1, \ell \rangle$ with $\ell=0, 1, 2$ and by a case of a
multilayer composite, present in both species.  Looking at
multilayered fiber in {\it B. subtilis} we observe synchronization of
the fiber and it's regulators. Since multi-layered building block in
bacillus is fully synchronized, we predict that there should be a AR
on the regulator SpoIIID that will turn this in a $\rvert 0, 1 \rangle$ fiber
in order to explain this synchronization.  Finding these missing links
is a useful byproduct of the existence of symmetries, which can be
done systematically to find and annotate new regulatory interactions.

The activity of the genes in operons are
reported individually in Ecomics, so we use the activity of the
individual genes (genes in one operon are marked in the plots).
Synchronization within operons is a trivial finding, and the test of
fiber synchronization is done by comparing the activity of any gene in
the operon with the genes outside the operon.
Moreover, the fiber predicts no synchronization between any gene in
the operon and the external regulator.

It is important to note the lack of synchronization between the fiber
genes {\it purR-pyrC} and its regulator {\it fur} as predicted by
fibrations.  This is despite the fact that the fiber is the regulon of
{\it fur}, that is, direct regulation does not lead to
synchronization.  As predicted, genes are highly coexpressed within
fibers, but not significantly correlated with the regulator genes.

Observed correlations largely confirm the synchronization within
fibers.
However, there are some interesting exceptions. For
instance, the genes {\it cssR} and {\it cssS} in {\it B. subtilis}
present large anticorrelations with a fiber (the one containing the
gene {\it cgeC}). This unexpected anti-correlation may indicate extra
transcriptional regulations between these fibers. These type of
correlations can be used to guide in the search for missing regulation
edges, which are ubiquitous in genetic network reconstructions.

\subsection{Coexpression of regulators of alternative carbon source
  utilization}
\label{sec:carbon}

\begin{figure*}
  \centering
  \includegraphics[width=0.8\linewidth]{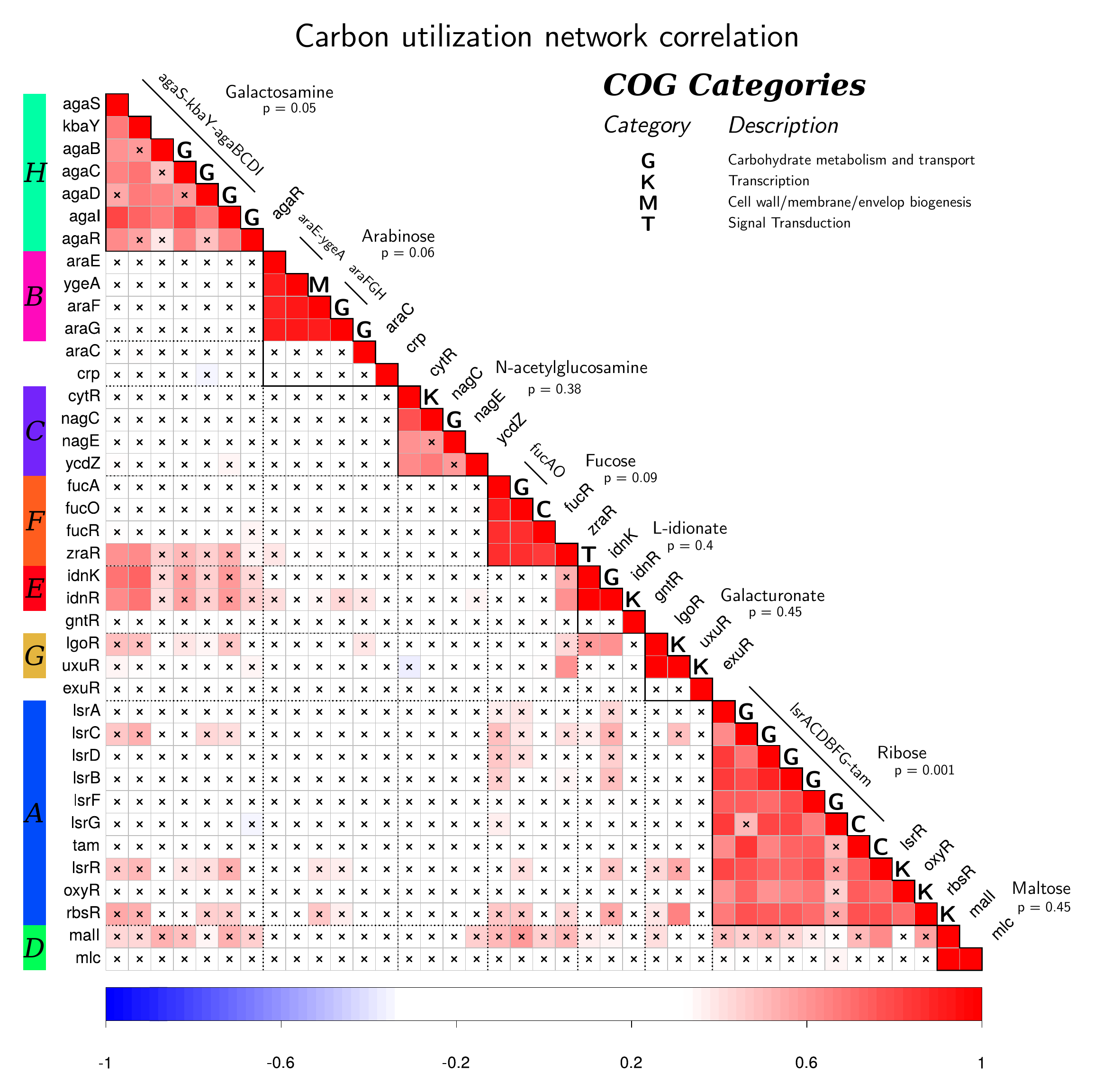}
  \caption{Carbon utilization: correlation matrix. Correlation matrix of the fiber building blocks involved in the carbon utilization system. Colored rectangles $A, B \dots H$ on the left code gene names that will be used in expression matrix plot Fig.~\ref{carbon0} and structure vs function plot Fig.~\ref{carbon2}. Operons are shown with lines along the correlation matrix diagonal. Black crosses show correlation entries below $0.6$ to compare low cross-correlation with high correlation inside fibers. COG categories are obtained using UniProt database \cite{uniprot}. Function of each block (Galactosamine, Arabinose, etc.) is defined by the type of it's regulator obtained from RegulonDB \cite{regulondb}.}
  \label{carbon1}
\end{figure*}

\begin{figure*}
  \centering
  \includegraphics[width=\linewidth]{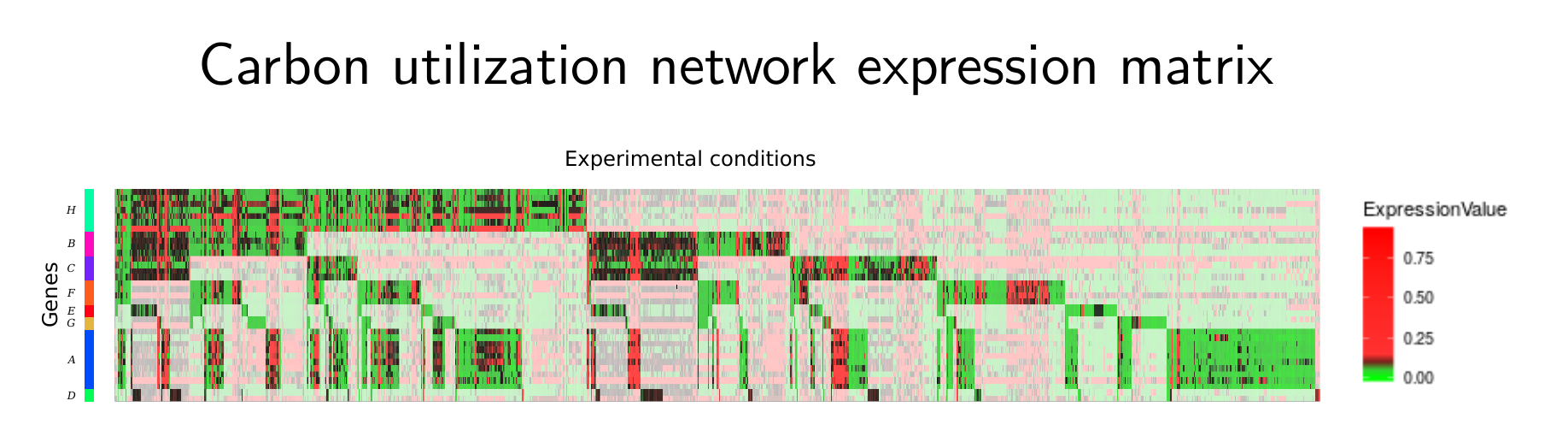}
  \caption{Carbon utilization: expression profile. Expression profile of 39 genes involved in the carbon utilization system over 1575 experimental conditions. Gene names correspond to the building blocks $A, B \dots H$ defined in Fig.~\ref{carbon1}. Conditions in white are filtered out using the method of ICV described in Section \ref{sec:MethodsICV} and the rest of the conditions are used to calculate correlation represented in Fig.~\ref{carbon1}.}
  \label{carbon0}
\end{figure*}

\begin{figure}
  \centering
  \includegraphics[width=\linewidth]{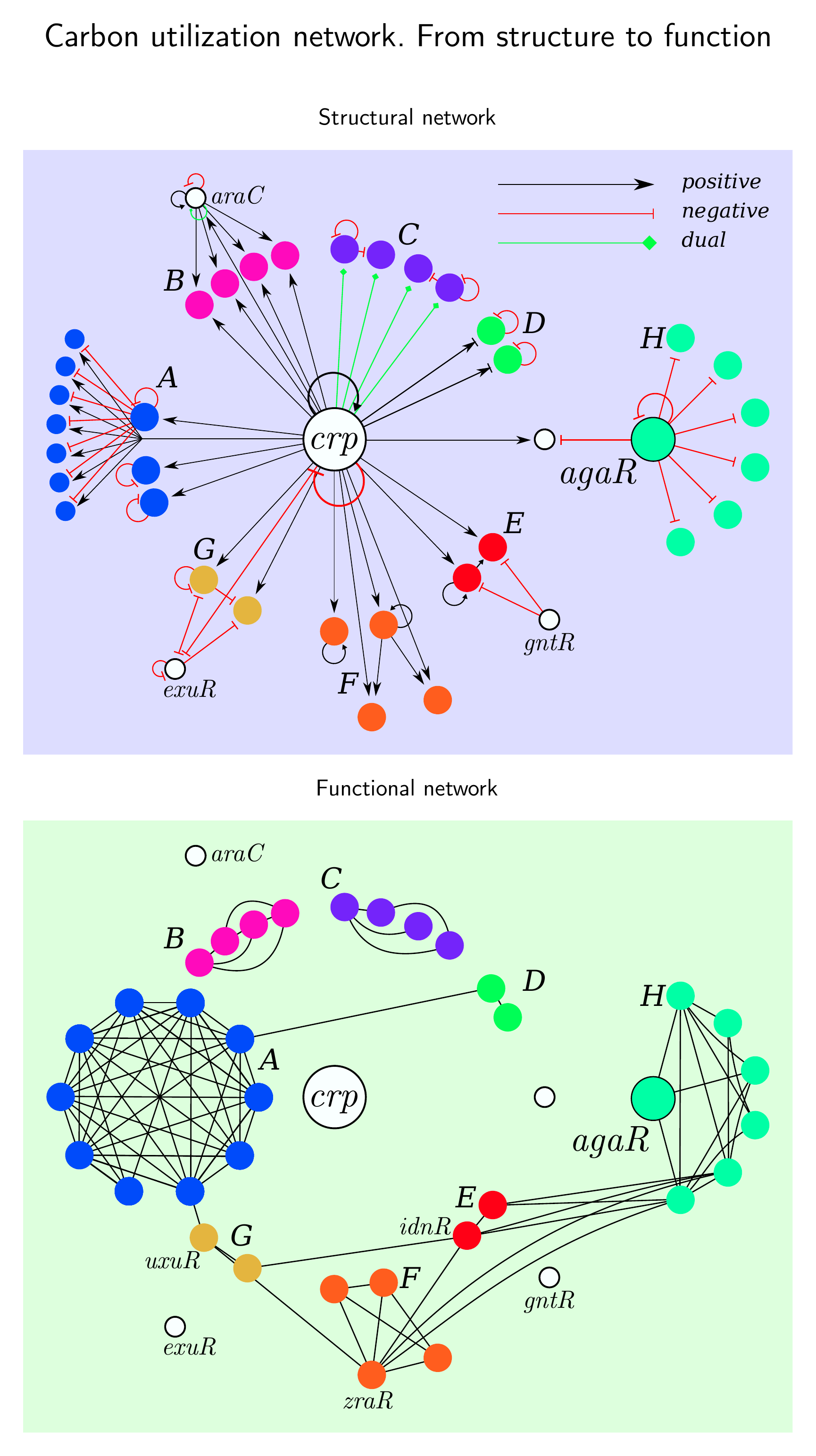}
  \caption{Carbon utilization: structural network vs functional network. Top part shows topology of the transcriptional regulation of the carbon utilization system. Bottom part shows the functional network of the carbon utilization system based on the Pearson correlation from Fig.~\ref{carbon1}. Correlation $C(i,j)$ is thresholded at $C(i,j) > 0.6$ to produce a functional network.}
  \label{carbon2}
\end{figure}

An underlying assumption in the functional annotation of genes is that
the genes that are highly connected in modules have a functional
relation among them; an assumption that is usually called 'guilty by
association' \cite{klipp,newman}. We now scrutinize this gene
annotation method in light of the existence of fibers.

For this purpose we take a closer look at a number of fibers involved
in the well studied functional module regulated by the master
regulator {\it crp} involved in the regulation of carbon source
catabolism, a well-characterized gene regulatory system. We will also
study how how the fibers are integrated into a larger network.  It has
previously been found that this circuit contains many types of network
motifs, including feed-forward loop (FFLs), FANs and others
\cite{alon,kaplan}. We will see that these network motifs are not
related to the synchronized fibers.

To build the carbon utilization network we select those building
blocks in which the TF (that generates the input tree) belongs to the
alternative carbon sources functional category \cite{alon-carbon} that
catabolizes sugars in the absence of the main source of sugar intake
that is glucose. This includes the catabolism of a number of sugars:
maltose, ribose, galactosamine, fucose, N-acetylglucosamine,
L-idonate, and galacturonate.  The regulation network is shown in
Fig.~\ref{carbon1}. Gene expression profile corresponding to this network demonstrating filtering
explained in Section \ref{sec:MethodsICV} is shown in Fig.~\ref{carbon0}.

Even thought genes involved in carbon utilization have a related
common biological function, that is the assimilation of alternative
carbon sources, these genes may not be expressed all together. Indeed,
each unit of this group must be activated in the presence of the
effector involved in the regulation.  Different building blocks are
activated under different nutrient intakes of sugars like the presence
of arabinose, maltose or ribose.  Most of these systems are
additionally regulated by the master regulator {\it crp}.

We start the analysis with the TF AgaR that catabolizes Galactosamine
that is regulated by {\it crp}, this TF regulates the fiber {\it
  agaS-kbaY-agaBCDI}. AgaR has several binding sites on the downstream
region of AgaS. The binding sites repress the promoter agaSp. Also
agarR is only regulated by AgaR with two binding sites on the
downstream region of gene {\it agaR} near promoter agaRp.  From
literature \cite{kaplan} it is known that {\it araJ} and {\it araE}
code for low-affinity transporters, while the {\it araFGH} operon
codes for a high-affinity transporter, these transport systems differ
in other properties in addition to their affinity for their substrate,
that is, they have different physiological properties. The {\it
  araFGH} operon is subject to strong catabolic repression, that is,
it is not expressed if glucose is present in the medium. On the other
hand, the low-affinity transporter {\it AraE} works at moderate
arabinose concentrations.  The expression of this transporter favors
the entry of the substrate and the expression of the enzymes that are
going to metabolize.


We found that the genes in the alternative carbon utilization system
show measurable coexpression within fibers, just as
predicted (see Fig.~\ref{carbon1}).  The coexpression in the fibers
(e.g., malI, mlc with AraC) is captured by the fibration
symmetry. This result shows also that the share of input functions
among genes, like in a regulon/operon does not necessarily lead to
synchronization.

We have also found fibers that contain genes with different input
functions like the ones in ribose circuit, yet, they still have measurably high co-expression.  They
are also correlated by function, since NagE is a N-acetylglucosamine
PTS permease, while YcdZ is a putative transmembrane protein that have
been predicted to interact with several sugar PTS permeases
(Uniprot). Thus, fibers do not only hint at synchronization, but also
at putative functional relations.

We also observe many network motifs in the carbon network as it was
found previously
\cite{alon,smma:02,alon-ffl,alon-delay,alon-ar,alon-carbon}. For
instance the genes \textit{gntR}, \textit{idnR}, \textit{idnK} form a FFL as shown in
Fig.~\ref{carbon2}. However these network motifs are not necessarily
related to the functional fibers of synchronized blocks.

\section{Summary and outlook}

Their potential for implementing controllable gene coexpression and
for facilitating gene rearrangements make gene fibers an interesting
tool of analysis not only for studying existing GRNs but also for
synthetic biology. Our results show that gene fibers can capture measurable gain of
synchronization in gene expression in two
well-reconstructed genetic networks, those of \emph{E.~coli} and
\emph{B.~subtilis}.
With this experimental confirmation of our
symmetry hypothesis, fibration symmetry seems to be a plausible
starting point for a broader theory of gene synchronization. Such a
theory would start with the description of exact symmetries and would
then proceed with perturbative schemes, allowing for heterogeneities,
based on controlled loop expansions on the theory \cite{coram}.  The
overall goal would be a predictive framework for gene synchronization,
including an assessment of the effects of mutations.  In the present
study, we focused completely on wild-type strains. It would be
interesting to study knockout experiments, where fibrations might
predict a loss of synchronization in comparison to the wild-type
predictions.

\clearpage

\section{Methods}

\subsection{Network construction}
\label{sec:NetworkConstruction}

Gene regulatory network of \emph{E.~coli} was obtained from operon
dataset from RegulonDB \cite{regulondb} with additional filtering.  An
operon starts with a promoter, but can also have internal promoters
and terminators. In a study in \emph{E.~coli}, about 45\% of all genes
were found to be single genes, about 20\% were in traditional operons
with one promoter (plus 7\% for operons with several terminators), and
about 20\% were in operons with internal promoters (plus 8\% for
operons with internal terminators) \cite{conway}.  In our networks, we
consider each operon as a single node, unless an operon contains
several TF, in which case, the TFs are considered individually
separately, but the genes in the operons that do not express a TF,
like for instance enzymes, are considered as a single node together
with one of the TF in the operon. For instance, the operon {\it
  gadAXW} in {\it E. coli} is considered as the operon {\it gadAX}
which includes one TF, and the other TF, GadW. For detailed
description of filtering process in \emph{E.~coli} see \cite{pnas} (SI
Chapter III).

\emph{B.~subtilis} gene regulatory network was obtained from SubtiWiki
\cite{SubtiWiki} with additional filtering. All sigma factor genes
were removed from the network. Additionally all types of link like
"positive\_regulation", "transcription\_activation" and
"transcriptional\_activation" were assigned as "activation" and
"anti-activation", "auto-repression", "negative\_autoregulation",
"transcription\_repression", "autorepression" and
"negative\_regulation" as "repression".

\subsection{Gene fibrations}
\label{sec:Fibrations}

When two nodes in a directed graph have isomorphic input trees, the
nodes are symmetric and synchronize their activity even if they are
not in the orbit of an automosphism. In this case, the synchronized
nodes are said to belong to the same fiber \cite{pnas, ian,ian-theory,
  vigna, aldis}.  The symmetry fibration is then a transformation that
collapses nodes in a fiber into a single node called the base and thus
reduces the circuit to its most simple form.  The ``orbits'' of
synchronized genes in an automorphism are now called the nodes in the
fiber produced by the symmetry fibration.

Groups of nodes that share fibration symmetry are called fibers.
Nodes that have isomorphic input tree belong to the same fibers of
minimal fibrations (further referred as fibrations for simplicity)
\cite{vigna}.  The input trees of all the genes in the tryptophan
circuit are isomorphic. They consist of a infinite chain as shown in
Fig.~\ref{ar}b, since this circuit contains one single loop and no
external regulators that do not belong to the fiber. Thus, we
characterized it by the fiber numbers $\rvert 1, 0\rangle$.  The symmetry
fibration is a transformation that reduces this circuit by collapsing
all nodes in the fiber to one, called the base. This is only possible
since all genes in a fiber are redundant in a dynamical state.


\subsection{\textcolor{red}{Equivalence between fibers of symmetry fibration and minimal balanced coloring}}
\label{sec:EquivalenceFibers}

\textcolor{red}{Equivalence between fibers of symmetry (surjective minimal) fibration and minimal balanced coloring (or coarsest equitable partition) is formally proven in Chapter 4 in \cite{aldis}. In particular, Theorem 4.7 states that maximal balanced equivalence relation is equivalent to the isomorphism relation between input trees of the infinite depth. That is, minimal balanced coloring induced by the maximal balanced equivalence relation is equivalent to having classes of nodes with isomorphic input trees, which correspond to the fibers of the symmetry fibration. Rigorous proof requires a fairly involved mathematical analysis \cite{aldis}, so we will only give a brief idea here. Consider graph $G$ with fibers $f_1, f_2, \dots f_n$ and balanced coloring $C=\{c_1, c_2, \dots c_m\}$}.

\textcolor{red}{1. Let there be two nodes $n \in f_i$ and $m \in f_j$ that belong to different fibers and have the same color $c_k$. Since $n$ and $m$ are of the same color, they will have isomorphic input trees. Therefore, there exists a fibration $\varphi$ that can collapse $f_i$ and $f_j$. Hence, $\psi$ is not minimal, which contradicts the assumption.}

\textcolor{red}{2. Let there be a fiber $f_i$ that breaks into two colors $c_j$ and $c_k$. Since $C$ is minimal, nodes of the different colors will have input trees that are not isomorphic. Therefore, there is a node $n \in c_j$ and a node $m \in c_k$ input trees of which are not isomorphic. Therefore, $n$ and $m$ can't belong to the same fiber, as required.}

\textcolor{red}{Consequently, there can't be any two nodes that belong to the same fiber, but have different colors and the opposite. Ergo, fibers of symmetry fibration are equvalent to minimal balanced coloring.}

\subsection{Algorithm for balanced coloring to identify fibers}
\label{sec:AlgorithmBalancedColoring}

Several algorithms can be used to find fibers in networks
\cite{CardonCrochemore,kamei,aldis}. All available algorithms are
based on finding 'balanced equivalence' relations in the network, see
\cite{golubitsky} for details. Current algorithms are based on the
algorithm introduced in 1982 by Cardon \& Crochemore
\cite{CardonCrochemore}. In \cite{pnas,ian} we have used the version
developed by Kamei \& Cock \cite{kamei}. A detailed explanation of
this algorithm is given in \cite{pnas,ian}. In a recent review article
we further discuss a fast algorithm that is scalable to large system
sizes in \cite{higor}.  The code of the algorithm in R can be accessed
at \url{https://github.com/ianleifer/fibrationSymmetries}.

\subsection{Gene expression data}
\label{sec:MethodsDataSets}

Expression data for \emph{E.~coli} were obtained from Ecomics
\cite{ecomics} (Multi-Omics Compendium for {\it E.~coli}).  Ecomics
contains microarray and RNA-seq experiments gathered from NCBI Gene
Expression Omnibus (GEO) \cite{geo}, for several \textit{E.~coli}
strains in different experimental growth conditions 1575 for 4096
genes.  We only used data from WT strains. The advantage of Ecomics
datasets compared with others compilations of expression experiments
like Colombos \cite{colombos} is that they provide absolute expression
levels instead of fold-changes.  Expression data for
\emph{B.~subtilis}) were obtained from SubtiWiki \cite{SubtiWiki}.
Subtwiki contains data from the GSE27219 experiment that has 104
experimental conditions for genes in wild type \emph{B.~subtilis}.

Using these data for a global analysis would be difficult since they
stem from different platforms users by the different experimental
groups.  Thus, raw data on gene expression among different experiments
from different labs is pre-processed by the curators of Ecomics and
SubtiWiki to produce normalized expression levels across platforms and
experiments by using noise reduction and bias correction normalized
data across different platforms. For our analysis, we selected data
from wild-type strains only (selecting WT conditions in 'strain',
'medium' and 'stress') to ensure the behavior of genes on standard
growth conditions without genotype modification from gene
knockouts. The 'perturbation' conditions in the Ecomics dataset
referring to mutants strains were not taken into account, and we use
always the same {\it E. coli} strain.


\subsection{Data and code availability}
\label{sec:DataAvailability}

The datasets used in this study are available at
Refs. \cite{regulondb} and \cite{SubtiWiki} and code for fiber finder
used in this study can be downloaded at
\url{https://github.com/MakseLab}.

\subsection{Selecting relevant experimental data based on the
  Inverse Coefficient of Variation}
\label{sec:MethodsICV}

To select experimental samples in which a gene set of interest is
active, ie, significantly expressed above random noise level, we used
the Inverse Coefficient of Variation (ICV) as a criterion similar to
the approach used by Colombos \cite{colombos}. We consider the genes
in the fiber and obtain the expression levels for all conditions for
the genes.  Then we calculate ICV for all conditions using the
following equation as is done in Colombos (details on the expression
analysis can be found at Ref. \cite{colombos} and at
\url{https://doi.org/10.1371/journal.pone.0020938.s001}, see Methods
Section \ref{sec:MethodsICV}.

\begin{equation}
  ICV_t=\dfrac{{\mu_t}}{\sigma_t},
\end{equation}

where $\mu_t$ is the average expression level of the chosen genes of
the fiber in the condition $t$ and $\sigma_t$ is the standard
deviation.  Following \cite{colombos}, we select conditions with
$ICV_t> <ICV_t>$, i.e., where the average expression levels in the particular
condition $t$ are higher than certain threshold that is given by the average ICV for all conditions of the fiber.  This score
reflects the fact that, in a relevant condition, the genes show an
increment on their expression above the individual variations caused
by random noise. ICV is a measure of scattering of the data. The more
scattered the data is compared with it's mean, the less is the value
of ICV.

We calculate the p-value of condition $t$ for the fiber genes using
z-score of the ICV coefficient for the selected condition using

\begin{equation}
  z_t = \frac{ICV_t-\mu_{ICV}}{\sigma_{ICV}},
\end{equation}

where,

\begin{eqnarray}
  \mu_{ICV} = <ICV_t>, \\
  \sigma_{ICV} = <ICV_t^2> - <ICV_t>^2
\end{eqnarray}

\section{Declarations}

Ethics approval and consent to participate: Not applicable.

Consent for publication: Not applicable.

Availability of data and materials. R package to reproduce the building blocks in \emph{E.~coli} and \emph{B.~subtilis} is available at \url{https://github.com/ianleifer/fibrationSymmetries} and \url{https://github.com/makselab}. Expression data for \emph{E.~coli} is available at \url{https://lipc23.engr.ccny.cuny.edu/f/636bca4a99f9440682ae/?dl=1} and for \emph{B.~subtilis} at \url{https://lipc23.engr.ccny.cuny.edu/f/10d13eb2f2e549b092a5/?dl=1}.

Competing interests: The authors declare no competing interest.

Funding: This work was funded by NIH-NIBIB  Grant No R01EB028157 and NIH-NIBIB Grant No R01EB022720 through the BRAIN Initiative.

Authors' contributions. I.L. designed research, performed research, curated data and wrote the paper. M.S.-P. designed research, performed research, curated data and wrote the paper. C.I. curated data and wrote the paper. H.M. designed research, performed research, curated data and wrote the paper.

Acknowledgements: We are grateful to W. Liebermeister for discussions.





\clearpage

\section{References}

\begin{thebibliography}
{\bibliographystyle}

\bibitem{monod} Monod J, Jacob F. General conclusions: teleonomic
  mechanisms in cellular metabolism, growth and differentiation.Cold
  Spring Harb Symp Quant Biol. 1961;26:389‐401

\bibitem{klipp} Klipp E, Liebermeister W, Wierling C, Kowald A, Herwig R. Systems Biology: a textbook. Weinheim: Wiley-VCH; 2016.
  
\bibitem{palsson} Palsson B\O. Systems biology: properties of reconstructed networks. (Cambridge University Press, New York, 2006).

\bibitem{gerosa} Gerosa L, Kochanowski K, Heinemann M, Sauer U. Dissecting specific and global transcriptional regulation of bacterial gene expression. Mol Syst Biol. 2013;9:658. Published 2013 Apr 16.

\bibitem{karlebach} Karlebach G, Shamir R. Modelling and analysis of gene regulatory networks. Nat Rev Mol Cell Biol. 2008;9(10):770‐780.

\bibitem{caldarelli} Buchanan M, Caldarelli G, De Los Rios P, Rao F, Vendruscolo M (editors). Networks in Cell Biology. (Cambridge University Press, Cambridge, 2010).

\bibitem{young} Lee TI, Rinaldi NJ, Robert F, Odom DT, Bar-Joseph Z, Gerber GK, Hannett NM, Harbison CT, Thompson CM, Simon I, Zeitlinger J, Jennings EG, Murray HL, Gordon DB, Ren B, Wyrick JJ, Tagne JB, Volkert TL, Fraenkel E, Gifford DK, Young RA. Transcriptional regulatory networks in Saccharomyces cerevisiae. Science. 2002;298(5594):799‐804.

\bibitem{arenas} Arenas A, Diaz-Guilera A, Kurths J, Moreno Y, Zhou C. Synchronization in complex networks. Physics Reports. 2008 Dec;469(3):93-153.
  
\bibitem{cecilia1} Lenz P, S{\o}gaard-Andersen L. Temporal and spatial oscillations in bacteria. Nat Rev Microbiol. 2011;9(8):565‐577. Published 2011 Aug 15.

\bibitem{cecilia2} Mir\'o-Bueno JM, Rodríguez-Pat\'on A. A simple negative interaction in the positive transcriptional feedback of a single gene is sufficient to produce reliable oscillations. PLoS One. 2011;6(11):e27414.
  
\bibitem{clocks} Fussenegger M. Synthetic biology: Synchronized bacterial clocks. Nature. 2010;463(7279):301‐302.

\bibitem{hasty} Hasty J, Dolnik M, Rottsch\"{a}fer V, Collins JJ. Synthetic gene network for entraining and amplifying cellular oscillations. Phys Rev Lett. 2002;88(14):148101. doi:10.1103/PhysRevLett.88.148101

\bibitem{stricker} Stricker J, Cookson S, Bennett MR, Mather WH, Tsimring LS, Hasty J. A fast, robust and tunable synthetic gene oscillator. Nature. 2008;456(7221):516‐519.

\bibitem{tigges} Tigges M, Marquez-Lago TT, Stelling J, Fussenegger M. A tunable synthetic mammalian oscillator. Nature. 2009;457(7227):309‐312.

\bibitem{alon} Alon U. An Introduction to Systems Biology: Design Principles of Biological Circuits. Boca Raton: CRC Press; 2006.

\bibitem{majt:08} Martinez-Antonio A, Janga SC, Thieffry D. Functional organisation of Escherichia coli transcriptional regulatory network. J Mol Biol. 2008;381(1):238‐247.

\bibitem{smma:02} Shen-Orr SS, Milo R, Mangan S, Alon U. Network motifs in the transcriptional regulation network of Escherichia coli. Nature Genet. 2002;31: 64-68.


\bibitem{gbmn:08} Goelzer A, Bekkal Brikci F, Martin-Verstraete I, Noirot P, Bessi\`eres P, Aymerich S, Fromion V. Reconstruction and analysis of the genetic and metabolic regulatory networks of the central metabolism of Bacillus subtilis. BMC Syst Biol. 2008;2:20. Published 2008 Feb 26.

\bibitem{alon-motif} Milo R, Shen-Orr SS, Itzkovitz S, Kashtan N, Chklovskii D, Alon U. Network motifs: simple building blocks of complex networks. Science 2002;298: 824-827.

\bibitem{pnas} Morone F, Leifer I, Makse HA. Fibration symmetries uncover the building blocks of biological networks. Proc Natl Acad Sci USA. 2020;117(15):8306‐-8314

\bibitem{connectomeMorone} Morone F, Makse HA. Symmetry group factorization reveals the structure-function relation in the neural connectome of {\it Caenorhabditis elegans}. Nat Commun. 2019; 10(1):4961. Published 2019 Oct 31.

\bibitem{ian} Leifer I, Morone F, Reis SDS, Andrade JS, Sigman M, Makse HA. Circuits with broken fibration symmetries perform core logic computations in biological networks. PLoS Comput Biol 2020: 16(6): e1007776.

\bibitem{grothendieck} Grothendieck A. Technique de descente et th\'eor\'emes d'existence en g\'eom\'etrie alg\'ebrique, I. G\'en\'eralit\'es. Descente par morphismes fid\'element plats. S\'eminaire N. Bourbaki {\bf 5}, Talk no. 190, p. 299-327 (1958-1960).
	
\bibitem{vigna} Boldi P, Vigna S. Fibrations of graphs. Discrete Mathematics. 2001;243: 21-66.
	
\bibitem{golubitsky} Golubitsky M, Stewart I. Nonlinear dynamics of networks: the groupoid formalism. Bull Am Math Soc. 2006;43: 305-364

\bibitem{wgcna}
  Horvath S (2011). Weighted Network Analysis: Application in Genomics and Systems Biology. New York, NY: Springer.

\bibitem{zhang-horvath} Zhang B, Horvath S (2005). "A general
  framework for weighted gene co-expression network analysis"
 Statistical Applications in Genetics and Molecular
  Biology. 4: 17.
  
\bibitem{r-package}
  Langfelder P, Horvath S (29 December 2008). "WGCNA: an R package for weighted correlation network analysis". BMC Bioinformatics. 9: 559.

\bibitem{wisdom} Marbach et al. Wisdom of crowds for robust gene
  network inference Nat Methods. ; 9(8):
  796–804. doi:10.1038/nmeth.2016.
  
\bibitem{bansal} Bansal,M. et al. (2007) How to infer gene networks
  from expression profiles. Mol.  Syst. Biol., 3, 78.

\bibitem{brugere}
  Network Structure Inference, A Survey: Motivations, Methods, and Applications
IVAN BRUGERE, University of Illinois at Chicago, USA
BRIAN GALLAGHER, Lawrence Livermore National Laboratory, USA TANYA Y. BERGER-WOLF

\bibitem{atul}
  Atul J. Butte and Isaac S. Kohane. 2000. Mutual Information Relevance Networks: Functional Genomic Clustering UsingPairwiseEntropyMeasurements.Paci cSymposiumonBiocomputing5(2000),415–426. 
  
\bibitem{hartung} Hartung, Thomas; Kleensang, Andre; Tran, Vy;
  Maertens, Alexandra (2018). "Weighted Gene Correlation Network
  Analysis (WGCNA) Reveals Novel Transcription Factors Associated With
  Bisphenol A Dose-Response". Frontiers in Genetics. 9: 508.

\bibitem{chen} Chen Y, Zhu J, Lum PY, Yang X, Pinto S, MacNeil DJ,
  Zhang C, Lamb J, Edwards S, Sieberts SK, Leonardson A, Castellini
  LW, Wang S, Champy MF, Zhang B, Emilsson V, Doss S, Ghazalpour A,
  Horvath S, Drake TA, Lusis AJ, Schadt EE (27 March
  2008). "Variations in DNA elucidate molecular networks that cause
  disease". Nature. 452 (7186): 429–35.

\bibitem{roy} Roy et al. BMC Bioinformatics 2014, 15(Suppl 7):S10
  Reconstruction of gene coexpression network from microarray data
  using local expression patterns Swarup Roy, Dhruba K Bhattacharyya,
  Jugal K Kalita.


\bibitem{collins}
Reverse engineering gene networks: Integrating genetic perturbations with dynamical modeling
Jesper Tegnér, M. K. Stephen Yeung, Jeff Hasty, and James J. Collins
PNAS May 13, 2003 100 (10) 5944-5949

\bibitem{wolf-network} Brugere I, Gallagher B, Berger-Wolf T. Network Structure Inference, A Survey: Motivations, Methods, and Applications. ACM Computing Surveys. 2016;51(2).

\bibitem{kaplan} Kaplan S, Bren A, Zaslaver A, Dekel E, Alon U. Diverse two-dimensional input functions control bacterial sugar genes. Mol Cell. 2008;29(6):786‐792.

\bibitem{nca} Liao JC, Boscolo R, Yang YL, Tran LM, Sabatti C, Roychowdhury VP. Network component analysis: reconstruction of regulatory signals in biological systems. Proc Natl Acad Sci USA. 2003;100(26):15522‐15527.

\bibitem{newman} Girvan M, Newman ME. Community structure in social and biological networks. Proc Natl Acad Sci U S A. 2002;99(12):7821‐7826.

\bibitem{ian-theory} Leifer I, M\'akse HA. Discovering synchrony in information-processing networks using graph fibrations. In preparation (2020).


\bibitem{sorrentino2016b} \textcolor{red}{Sorrentino F, Pecora L. Approximate cluster synchronization in networks with symmetries and parameter mismatches. Chaos. 2016 Sep;26(9):094823.}

\bibitem{deville} DeVille L, Lerman E. Modular dynamical systems on networks. J. Eur. Math. Soc. 17, 2977–3013 (2015).

    
\bibitem{sanders} Nijholt E, Rink B, Sanders J. Graph fibrations and symmetries of network dynamics. J. Differ. Equ. 261, 4861–4896 (2016).

%
  
\bibitem{pecora1} Pecora LM, Sorrentino F, Hagerstrom AM, Murphy TE, Roy R. Cluster synchronization and isolated desynchronization in complex networks with symmetries. Nat Commun. 2014;5:4079. Published 2014 Jun 13.

\bibitem{pecora2} Sorrentino F, Pecora LM, Hagerstrom AM, Murphy TE, Roy R. Complete characterization of the stability of cluster synchronization in complex dynamical networks. Sci Adv. 2016;2(4):e1501737. Published 2016 Apr 22.

\bibitem{sorrentino2019} \textcolor{red}{Blaha K, Huang K, Della Rossa F, Pecora L, Hossein-Zadeh M and Sorrentino F. Cluster Synchronization in Multilayer Networks: A Fully Analog Experiment with LC Oscillators with Physically Dissimilar Coupling. Physical Review Letters, 2019:122(1).}
  
\bibitem{sorrentino2020} \textcolor{red}{Della Rossa F, Pecora L, Blaha K, Shirin A, Klickstein I and Sorrentino F. Symmetries and cluster synchronization in multilayer networks. Nature communications. 2020:11(3179).}

\bibitem{kamei} Kamei H, Cock PJ. A. Computation of balanced equivalence relations and their lattice for a coupled cell network. SIAM J Appl Dyn Syst. 2013;12: 352-382.

\bibitem{regulondb} Gama-Castro S, Salgado H, Santos-Zavaleta A, Ledezma-Tejeida D, Muniz-Rascado L, Garcia-Sotelo JS, Alquicira-Hern\'andez K, Martinez-Flores I, Pannier L, Castro-Mondrag\'on JA, Medina-Rivera A, Solano-Lira H, Bonavides-Martínez C, P\'erez-Rueda E, Alquicira-Hern\'andez S, Porr\'on-Sotelo L, L\'opez-Fuentes A, Hern\'andez-Koutoucheva A, Del Moral-Ch\'avez V, Rinaldi F, Collado-Vides J. RegulonDB version 9.0: high-level integration of gene regulation, coexpression, motif clustering and beyond. Nucleic Acids Res. 2016;44(D1):D133‐D143.

\bibitem{SubtiWiki} Zhu B, St\"ulke J. SubtiWiki in 2018: from genes and proteins to functional network annotation of the model organism Bacillus subtilis. Nucleic Acids Res. 2018;46(D1):D743‐D748.

\bibitem{other0} Abrams DM, Pecora LM, Motter AE. Introduction to focus issue: Patterns of network synchronization. Chaos. 2016;26(9):094601.
  
\bibitem{grouptheory} Hamermesh H. Group Theory and its Application to Physical Problems. (Dover, New York, 1989).



\bibitem{pmid3106331} Grove CL, Gunsalus RP. Regulation of the aroH operon of Escherichia coli by the tryptophan repressor. J Bacteriol. 1987;169(5):2158‐2164.

\bibitem{Karp:2018aa} Karp PD, Ong WK, Paley S, Billington R, Caspi R, Fulcher C, Kothari A, Krummenacker M, Latendresse M, Midford PE, Subhraveti P, Gama-Castro S, Muniz-Rascado L, Bonavides-Martinez C, Santos-Zavaleta A, Mackie A, Collado-Vides J, Keseler IM, Paulsen I. The EcoCyc Database. EcoSal Plus. 2018;8(1):10.1128/ecosalplus.ESP-0006-2018.

\bibitem{luis} Alvarez LA, Leifer I, Liebermeister W, Makse HA. Identifying the minimal transcriptional regulatory network in bacteria. In preparation (2020).

\bibitem{common_inputs} Wang YX, Huang H. Review on statistical methods for gene network reconstruction using expression data. J Theor Biol. 2014 Dec 7;362:53-61.

\bibitem{bintu} Bintu L, Buchler NE, Garcia HG, Gerland U, Hwa T, Kondev J, Phillips R. Transcriptional regulation by the numbers: models. Curr Opin Genet Dev. 2005;15(2):116‐124.

\bibitem{hwa} Buchler NE, Gerland U, Hwa T. On schemes of combinatorial transcription logic. Proc Natl Acad Sci USA. 2003;100(9):5136‐5141.

\bibitem{mssz:06} Mayo AE, Setty Y, Shavit S, Zaslaver A, Alon U. Plasticity of the cis-regulatory input function of a gene. PLoS Biol. 2006;4(4):e45.

\bibitem{ecomics} Kim M, Rai N, Zorraquino V, Tagkopoulos I. Multi-omics integration accurately predicts cellular state in unexplored conditions for Escherichia coli. Nat Commun. 2016;7:13090. Published 2016 Oct 7.

\bibitem{geo} Barrett T, Wilhite SE, Ledoux P, Evangelista C, Kim IF, Tomashevsky M, Marshall KA, Phillippy KH, Sherman PM, Holko M, Yefanov A, Lee H, Zhang N, Robertson CL, Serova N, Davis S, Soboleva A. NCBI GEO: archive for functional genomics data sets--update. Nucleic Acids Res. 2013;41(Database issue):D991‐D995.

\bibitem{array} Kolesnikov N, Hastings E, Keays M, Melnichuk O, Tang YA, Williams E, Dylag M, Kurbatova N, Brandizi M, Burdett T, Megy K, Pilicheva E, Rustici G, Tikhonov A, Parkinson H, Petryszak R, Sarkans U, Brazma A. ArrayExpress update--simplifying data submissions. Nucleic Acids Res. 2015;43(Database issue):D1113‐D1116.

\bibitem{colombos} Moretto M, Sonego P, Dierckxsens N, Brilli M, Bianco L, Ledezma-Tejeida D, Gama-Castro S, Galardini M, Romualdi C, Laukens K, Collado-Vides J, Meysman P, Engelen K. COLOMBOS v3.0: leveraging gene expression compendia for cross-species analyses. Nucleic Acids Res. 2016;44(D1):D620‐D623.

\bibitem{cramerStatistics} Cramer H. Mathematical methods of
  statistics, 9th edition. (GLS press, Bombay, 1961).

\bibitem{uniprot} The UniProt Consortium, UniProt: a worldwide hub of protein knowledge, Nucleic Acids Research, Volume 47, Issue D1, 08 January 2019, Pages D506–D515.

\bibitem{alon-carbon} Aidelberg G, Towbin BD, Rothschild D, Dekel E, Bren A, Alon U. Hierarchy of non-glucose sugars in Escherichia coli. BMC Syst Biol. 2014;8:133. Published 2014 Dec 24.

\bibitem{alon-ffl} Mangan S, Alon U. Structure and function of the feed-forward loop network motif. Proc Natl Acad Sci USA. 2003;100(21):11980‐11985.

\bibitem{alon-delay} Mangan S, Zaslaver A, Alon U. The coherent feedforward loop serves as a sign-sensitive delay element in transcription networks. J Mol Biol. 2003;334(2):197‐204.
  
\bibitem{alon-ar} Madar D, Dekel E, Bren A, Alon U. Negative auto-regulation increases the input dynamic-range of the arabinose system of Escherichia coli. BMC Syst Biol. 2011;5:111. Published 2011 Jul 12.

\bibitem{leibler} Hartwell L, Hopfield J, Leibler S, Murray AW. From molecular to modular cell biology. Nature 1999:402, C47–C52.
  
\bibitem{module1} Brugere I, Gallagher B, Berger-Wolf TY. Network Structure Inference, A Survey: Motivations, Methods, and Applications. ACM Comput. Surv. 51, 2, Article 24 (April 2018).

\bibitem{module2} Pratapa A, Jalihal AP, Law JN, Bharadwaj A, Murali TM. Benchmarking algorithms for gene regulatory network inference from single-cell transcriptomic data. Nat Methods. 2020;17(2):147‐154.

\bibitem{module3} Marbach D, Costello JC, K\"uffner R, Vega NM, Prill RJ, Camacho DM, Allison KR, DREAM5 Consortium, Kellis M, Collins JJ, Stolovitzky G. Wisdom of crowds for robust gene network inference. Nat Methods. 2012;9(8):796‐804. Published 2012 Jul 15.

\bibitem{louvain} Blondel VD, Guillaume J-L, Lambiotte R, Lefebvre E. Fast Unfolding of Communities in Large Networks. J. Stat. Mech.: Theory and Experiment 10, P10008 (2008)

\bibitem{lasso} Friedman J, Hastie T, Tibshirani R. Sparse inverse covariance estimation with the graphical lasso. Biostatistics. 2008;9(3):432‐441.

\bibitem{bialek} Lezon TR, Banavar JR, Cieplak M, Maritan A, Fedoroff NV. Using the principle of entropy maximization to infer genetic interaction networks from gene expression patterns. Proc Natl Acad Sci U S A. 2006 Dec 12;103(50):19033-8.

  \bibitem{uber} Trends Biochem Sci. 2000 Oct;25(10):474-9.  Gene
    context conservation of a higher order than operons.  Lathe WC
    3rd1, Snel B, Bork P.

\bibitem{juri:16} Junier I, Rivoire O. Conserved Units of
  Co-Expression in Bacterial Genomes: An Evolutionary Insight into
  Transcriptional Regulation. PLoS One. 2016;11(5):e0155740. Published
  2016 May 19.


\bibitem{jacob-tinkerer} Jacob F. Evolution and tinkering. Science. 1977;196(4295):1161‐1166.

\bibitem{alon-tinkerer} Alon U. Biological networks: the tinkerer as an engineer. Science. 2003;301(5641):1866‐1867.

\bibitem{coram} Coram DS, Duval PF. Approximate fibrations. Rocky Mountain J. Math. 1977:7(2):275--288. 

\bibitem{pmte:99} Pellegrini M, Marcotte EM, Thompson MJ, Eisenberg D, Yeates TO. Assigning protein functions by comparative analysis: protein phylogenetic profiles. Proc. Natl. Acad. Sci. USA 1999;96(8):4285-8.

\bibitem{babg:14} Babu M, Arnold R, Bundalovic-Torma C, Gagarinova A, Wong KS, Kumar A, Stewart G, Samanfar B, Aoki H, Wagih O, Vlasblom J, Phanse S, Lad K, Yeou Hsiung Yu A, Graham C, Jin K, Brown E, Golshani A, Kim P, Moreno-Hagelsieb G, Greenblatt J, Houry WA, Parkinson J, Emili A. Quantitative genome-wide genetic interaction screens reveal global epistatic relationships of protein complexes in Escherichia coli. PLoS Genet. 2014;10(2):e1004120. Published 2014 Feb 20.

\bibitem{conway} Conway T, Creecy JP, Maddox SM, Grissom JE, Conkle TL, Shadid TM, Teramoto J, San Miguel P, Shimada T, Ishihama A, Mori H, Wanner BL. Unprecedented high-resolution view of bacterial operon architecture revealed by RNA sequencing. mBio. 2014;5(4):e01442-14. Published 2014 Jul 8.

\bibitem{aldis} Aldis JW. A polynomial time algorithm to determine maximal balanced equivalence relations. Int J Bifurc Chaos Appl Sci Eng. 2008;18: 407-427.

\bibitem{CardonCrochemore} Cardon A, Crochemore M. Partitioning a graph in O($\mid A \mid \log_2 \mid V \mid$). Theor. Comput. Sci. 1982; 19, 85–98.

\bibitem{higor} Leifer I, Monteiro HS, Reis S, Andrade JS, M\'akse HA. Fast and scalable algorithm to find fibers via symmetry fibrations in large scale networks. In preparation (2020).

\bibitem{zabl:13} Zelcbuch L, Antonovsky N, Bar-Even A, Levin-Karp A, Barenholz U, Dayagi M, Liebermeister W, Flamholz A, Noor E, Amram S, Brandis A, Bareia T, Yofe I, Jubran H, Milo R. Spanning high-dimensional expression space using ribosome-binding site combinatorics. Nucleic Acids Res. 2013;41(9):e98.

\bibitem{pssm} Hertz GZ, Hartzell GW 3rd, Stormo GD. Identification of consensus patterns in unaligned DNA sequences known to be functionally related. Comput Appl Biosci. 1990;6(2):81‐92.

\bibitem{logo} Medina-Rivera A, Abreu-Goodger C, Thomas-Chollier M, Salgado H, Collado-Vides J, van Helden J. Theoretical and empirical quality assessment of transcription factor-binding motifs. Nucleic Acids Res. 2011;39(3):808‐824.

\bibitem{weblogo} \url{https://weblogo.berkeley.edu/logo.cgi}


%
%
%
%
%
%
%
%
%
%
%

\end{thebibliography}
\end{document}